\documentclass[lettersize,journal]{IEEEtran}
\usepackage{amsmath,amssymb,amsfonts}
\usepackage{algorithmic}
\usepackage{algorithm}
\usepackage{array}
\usepackage[caption=false,font=normalsize,labelfont=sf,textfont=sf]{subfig}
\usepackage{textcomp}
\usepackage{stfloats}
\usepackage{url}
\usepackage{verbatim}
\usepackage{cite}
\usepackage{xcolor}
\usepackage{booktabs}
\usepackage{xspace}
\usepackage{enumitem}
\usepackage{makecell}
\usepackage{soul}
\usepackage{tabularx}   
\usepackage{booktabs}   
\usepackage{enumitem}   

\usepackage[most]{tcolorbox}
\newtcolorbox{highlighted}{colback=yellow,coltext=black,breakable}
\newtcolorbox{highlighted_c}{colback=pink,coltext=black,breakable}
\usepackage{subfig}
\usepackage{graphicx}

\usepackage{xcolor}
\usepackage{soul} 
\sethlcolor{yellow!22}

\DeclareRobustCommand{\Name}{Flowcut\xspace}
\DeclareRobustCommand{\name}{Flowcut\xspace}
\DeclareRobustCommand{\Names}{Flowcuts\xspace}
\DeclareRobustCommand{\names}{flowcuts\xspace}

\soulregister\Name0
\soulregister\name0
\soulregister\Names0
\soulregister\names0
\soulregister\NameHL0
\soulregister\nameHL0
\soulregister\NamesHL0
\soulregister\namesHL0
\soulregister\markword1
\soulregister\xspace0
\soulregister\textbf1
\soulregister\emph1
\soulregister\texttt1
\soulregister\textit1
\soulregister\textsc1
\soulregister\cite7
\soulregister\ref7
\soulregister\eqref7

\newcommand{\pktdata}{{\includegraphics[scale=1,trim=0 1 0 0]{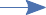}}}
\newcommand{\pktack}{{\includegraphics[scale=1,trim=0 2 0 0]{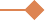}}}
\newcommand{\pktpause}{{\includegraphics[scale=1,trim=0 1 0 0]{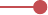}}}
\newcommand{\pktresume}{{\includegraphics[scale=1,trim=0 1 0 0]{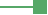}}}
\newcommand{\congestion}{{\includegraphics[scale=.4,trim=0 10 0 0]{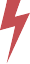}}}

\begin{document}

\title{Flowcut Switching: High-Performance Adaptive Routing with In-Order Delivery Guarantees}

\author{Tommaso Bonato,~\IEEEmembership{ETH Zurich}, Daniele De Sensi,~\IEEEmembership{Sapienza University of Rome}, Salvatore Di Girolamo,~\IEEEmembership{ETH Zurich}, Abdulla Bataineh,~\IEEEmembership{HPE}, David Hewson,~\IEEEmembership{HPE}, Duncan Roweth,~\IEEEmembership{HPE}, Torsten Hoefler,~\IEEEmembership{ETH Zurich}
}

\markboth{}%
{Shell \MakeLowercase{\textit{et al.}}: A Sample Article Using IEEEtran.cls for IEEE Journals}

\IEEEpubid{}

\maketitle

\begin{abstract}
Network latency severely impacts the performance of applications running on supercomputers. Adaptive routing algorithms route packets over different available paths to reduce latency and improve network utilization. However, if a switch routes packets belonging to the same network flow on different paths, they might arrive at the destination out-of-order due to differences in the latency of these paths. For some transport protocols like TCP, QUIC, and RoCE, out-of-order (OOO) packets might cause large performance drops or significantly increase CPU utilization. In this work, we propose \textit{Flowcut switching}, a new adaptive routing algorithm that provides high-performance in-order packet delivery. Differently from existing solutions like \textit{Flowlet switching}, which are based on the assumption of bursty traffic and that might still reorder packets, \textit{Flowcut switching} guarantees in-order delivery under any network conditions, and is effective also for non-bursty traffic, as it is often the case for RDMA. On top of this, Flowcut can be implemented either at the switch or NIC level providing flexibility and different tradeoffs.
\end{abstract}


\section{Introduction and Motivation}
\IEEEPARstart{T}{he} interconnection network is a performance critical component for applications running on data centers. With the emergence of new paradigms such as disaggregated memory~\cite{10.1145/3445814.3446713} and distributed storage~\cite{10.1145/3452296.3472940}, the network must guarantee high throughput and low latency, with close-to-zero queuing delay~\cite{199305}. 

Because most data center networks are characterized by multiple paths between any pair of endpoints~\cite{9248644,10.5555/3433701.3433736,vl2,conga}, \emph{adaptive routing} (a subset of \emph{traffic load balancing}) algorithms can be used to distribute packets on multiple paths, thus reducing queuing delays~\cite{loadbalancingsurvey}. In contrast, ECMP~\cite{ecmp} is a widely used routing algorithm, which distributes packets over equal-length paths based on the result of a hash function computed on some fields in the packet header. However, although widely adopted for its simplicity~\cite{10.1145/2829988.2787508}, vanilla ECMP does not use any congestion information when selecting paths. For this reason, packets often experience congestion, even when there are alternative non-congested paths that could have been used instead~\cite{10.1145/2535372.2535375,conga,drill,meta2024}. 
To avoid congestion, some routing algorithms estimate the congestion of different paths and route packets on the least congested path.
Routing decisions can be taken at different degrees of granularity, ranging from per packet~\cite{slingshot,drill} to per flow\footnote{A flow is usually defined as a sequence of packets from a source to a destination. For example, in an IP network a flow can consist of packets with the same 5-tuple, composed of: source and destination addresses, source and destination ports, and transport protocol.} basis~\cite{hedera,vl2,10.1145/1397718.1397732}, with some intermediate solutions~\cite{letitflow,flowcell}.

\subsection{Per Packet Adaptive Routing}
Adaptively routing on a per packet basis usually provides the lowest packet latency~\cite{loadbalancingsurvey}, because each packet could be sent on the least congested path~\cite{slingshot,6392830,6567015}. However, due to differences in latencies of different paths, packets of a flow can arrive at their destination in a different order. Because protocols like RoCE, TCP, and QUIC~\cite{10.1145/3405796.3405827,quic} need to deliver the packets to the application in-order, out-of-order (OOO) packets increase the flow completion time (FCT) due to reordering of the packets on the receiver side and/or packets retransmission~\cite{10.1145/3405796.3405827,Ghasemirahni2022PacketOM}. 

For example, it has been shown that, even if only 0.6\% of the packets are lost on a 40 Gbps network (and hence generate out-of-order packets), RoCE consumes 80\% of an Intel Xeon E5-2630 v3 2.4GHz core time~\cite{10.1145/3106989.3106993} to reorder those packets when using selective ACKs (i.e., almost one entire core is dedicated to packet reordering). 
Moreover, many RoCE implementations use the \textit{go-back-n} protocol to retransmit OOO packets~\cite{revisitingrdma,8895760}, further exacerbating the problem. This has been illustrated in a recent work \cite{left} where under per-packet multipath, the fraction of out-of-order packets skyrockets once load exceeds 0.3, and RDMA’s simple go-back-n retransmit drops goodput to below 10\% of ideal. Similarly, a recent study on the performance of several libraries implementing the QUIC protocol shows that, even when less than 1\% of the packets arrive out of order, this can cause a performance drop of up to 10 times if OOO packets are treated as packet lost, or a 30\% increase in the CPU usage if they are reordered~\cite{10.1145/3405796.3405827}.

Moreover, acceleration techniques like \textit{Generic Receive Offload} (GRO) in some cases are not applicable in presence of out-of-order packets, since they rely on in-order delivery~\cite{juggler} (e.g., Linux GRO~\cite{dpdk-gro}). Whereas reordering can be offloaded to programmable NICs to avoid wasting CPU cycles, this is usually a complex task and complicates the hardware design of these devices~\cite{10.1145/3405796.3405827}.  

\subsection{Per Flowlet/Flowcell Routing}
To reduce the number of OOO packets, some load routing algorithms take new decisions at a coarser granularity, for example, on a per \textit{Flowcell}~\cite{flowcell} or per \textit{Flowlet}~\cite{10.5555/3433701.3433736,letitflow,conga,sinha2004harnessing,10.1145/1232919.1232925} basis. A flowcell is a sequence of consecutive packets, with a cumulative length not larger than some threshold. A flowlet is instead defined as a sequence of consecutive packets with a variable cumulative length, and separated in time from the subsequent flowlet by a fixed time interval. Different flowlets can be routed on different paths and, if the time gap is sufficiently large, packets arrive at the destination in order.

\subsection{Limitations of Existing Algorithms}
Flowlet switching works well for bursty traffic~\cite{10.1145/1232919.1232925} (as it is often the case for TCP traffic), as burstiness creates more opportunities for the creation of new flowlets and, thus, for better balancing the traffic on different paths.
However, \textit{Remote Direct Memory Access} (RDMA) protocols such as \textit{RDMA over Converged Ethernet} (RoCE)~\cite{roce,revisitingrdma} and \textit{Scalable Reliable Datagram} (SRD)~\cite{9167399} are often used for east-west traffic in large data centers (like those of Microsoft~\cite{8895760,10.1145/2934872.2934908}, Google~\cite{278358,10.1145/2785956.2787510,10.1145/2785956.2787484,10.1145/3387514.3405897}, and Amazon~\cite{9167399}). Differently from TCP, on RDMA hardware rate limiters are often utilized~\cite{10.1145/2785956.2787484,10.1145/3267809.3267826,10.5555/3323234.3323245}. As a consequence, RDMA is often characterized by fewer flowlets~\cite{8895760,mprdma}, thus limiting adaptive routing opportunities when using Flowlet switching. 

Moreover, although algorithms based on flowcells and flowlets can reduce the number of out of order packets, they cannot completely guarantee in-order delivery. Indeed, in-order delivery can only be guaranteed if the threshold used to identify a burst (i.e., a flowlet), is larger than the latency difference between the paths on which the flowlets are forwarded. This depends on the workload running on the network, the network topology (e.g., when having paths of different lengths), and the underlying network conditions~\cite{benet2019flowdyn,letitflow}.

\begin{figure}[htpb]
\centering
\includegraphics[width=\columnwidth]{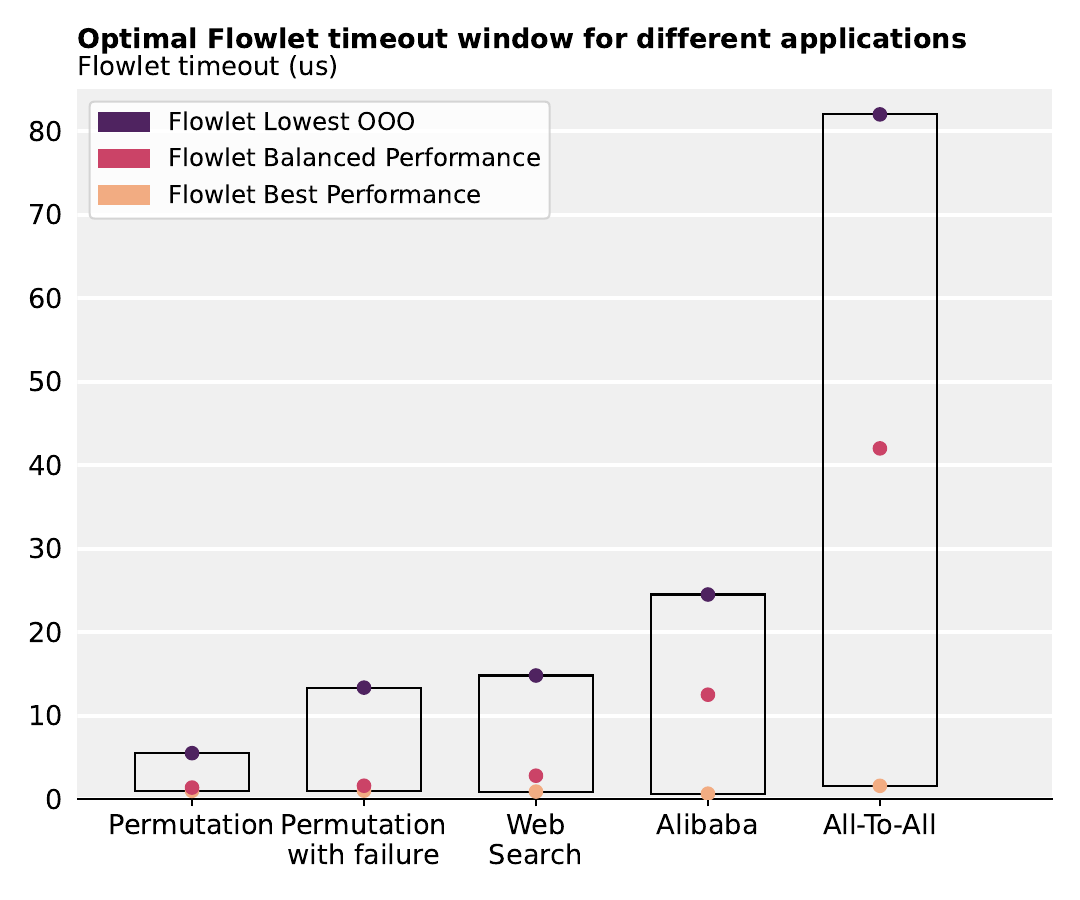}
\caption{Optimal flowlet window for different workloads and network conditions. The optimal performance is calculated using the average flow completion time. Web search and Alibaba are traces with their distribution presented in Figure~\ref{fig:distr}}
\label{fig:window}
\end{figure}

However, network conditions vary with time in unpredictable ways~\cite{10.1145/1879141.1879175,10.1145/1644893.1644918} due to, e.g., changes in traffic patterns and link failures~\cite{8057002}. Thus, having a fixed threshold for all the scenarios cannot be optimal. Figure~\ref{fig:window} shows how the best window size for Flowlet switching can heavily change depending on the type of workload being run. Moreover, even for the same workload, the optimal window can be affected by the input data and the actual network configuration. Finally, partial or total failures can dynamically change the network state and affect the optimal threshold choice (Figure~\ref{fig:window}).

While out of order delivery remains a challenge for many high‐performance transports, some recent NICs (e.g. NVIDIA Mellanox ConnectX-5 \cite{connectx5}) now include on-card reordering engines. However, these features are proprietary, tied to specific switch fabrics and transport protocols, and are not broadly available across vendors or workloads. By contrast, Flowcut’s in-flight-drain mechanism works entirely in software (or with minimal switch assistance), is transport-agnostic, and can be deployed on any commodity NIC without specialized hardware support.

\subsection{Flowcut Switching} \label{sec:flowcut_intro}
For the aforementioned reasons, in this paper we design a new adaptive routing algorithm for in-order RDMA traffic (RoCE), named \textit{\name switching}. Flowcut switching can adaptively route non-bursty traffic regardless of the network conditions, while providing at the same time in-order delivery guarantees. A \name is a sequence of consecutive packets of a flow, all being forwarded on the same path. Differently from Flowlet switching, each switch tracks the number of in-flight packets for each active flow, and creates a new \name (and route it on a different path) only if there are no in-flight packets for the flow between the switch and the destination host. By doing so, all the in-flight packets of a flow are always routed on the same path, thus guaranteeing in-order delivery. As an optimization to reduce memory footprint, Flowcut can also be implemented only at the ingress switch or the NIC level.

In addition, if the NIC or ingress switch (i.e., the switch directly connected to the source of the flow) detects that a flow is experiencing congestion and it cannot create a new \name (because there are still in-flight packets), it can temporarily pause the transmission of packets for that flow. When there are no in-flight packets for the flow, transmission is resumed, and a new \name is created. The new \name is routed on a different (and less congested) path, allowing non-bursty flows to react to congestion. 

The Rosetta switches of HPE's Slingshot network \cite{slingshot}, designed by some of the co-authors of this paper, implement key ideas that we discuss here, such as the use of flowcuts and an intelligent draining process. In our model, we use timing-based congestion detection to initiate re-routing, which is a key strategy implemented by Slingshot to detect and respond to congestion.
We evaluate \name switching through simulations on popular data center network topologies (blocking and non-blocking fat trees~\cite{9167399,10.1145/2829988.2787508}, and Dragonfly~\cite{278358}) and workloads, and we compare it with different state of the art adaptive routing algorithms. We show that \name switching improves \textit{flow completion time} (FCT) up to 50\% with respect to per flow routing algorithms like ECMP, and up to 40\% compared to Flowlet switching (when setting the timeout value to re-order only few packets, more details in Section~\ref{sec:fat_tree_results}), while guaranteeing in-order delivery of packets. Moreover, we also showcase how Flowcut switching can handle failures effectively and how it beats ECMP by 5x in such scenario.

\section{General Design}\label{sec:flowchip}

By design, \Name switching supports three deployment models to accommodate hardware and operational constraints:
\textbf{Full switch} (state and re-routing at every switch); \textbf{Ingress only} (state and re-routing only at the ingress switches); and
\textbf{NIC only} (state and re-routing at the endpoints). In Section~\ref{flowcut_variants_result}, we show that while the \textbf{Full switch} deployment achieves the best overall performance, the far more memory-efficient \textbf{NIC only} variant performs comparably in all tested scenarios.

We discuss in the following how Flowcut can be implemented in the switches, and in Section~\ref{sec:variant:nic} how it can be implemented in the NICs without any support from the switches. As previously mentioned, when using \name switching switches needs to keep a small state (a few bytes) for each active flow (i.e., for flows that have in-flight packets). The \name state includes the port on which the last packet for the flow has been transmitted, the port from which that packet has been received, and the number of in-flight bytes. Each switch stores information about active \names in a \textit{\name table}, indexed by a unique key (e.g., the 5-tuple for IP flows), and implemented, for example, as a \textit{content-addressable memory} (CAM). We show in Section~\ref{sec:evaluation:consumption:memory} that given current network characteristics, the \textit{\name table} can fit in the memory of most existing switches.

\begin{figure}[htpb]
\centering
\includegraphics[width=\columnwidth]{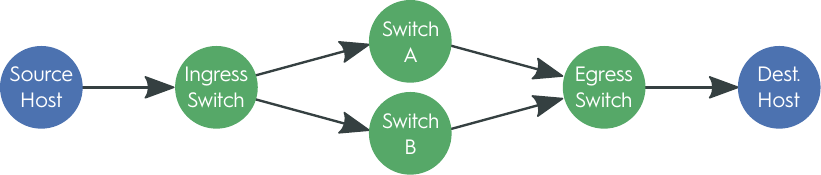}
\caption{Network example.}
\label{fig:flowcut_topology}
\end{figure}

To simplify the exposition, we show in Figure~\ref{fig:flowcut_topology} a simple reference topology. We define as \textit{ingress} switch the switch connected to the source host where the flow originates, and as \textit{egress} switch the one connected to the destination host of the flow. Although in this example there is a single hop between ingress and egress switches (traffic goes either through \textit{Switch A} or \textit{B}), \name switching works for any arbitrary number of hops.

\subsection{\Name Creation and Termination}
The switch keeps the information on a flow in the \name table as long as there are in-flight packets for that flow. When a switch receives a packet, it uses the flow key to index the \name table. If an entry is found, the packet is forwarded to the output port stored in the Flowcut table entry, and the number of in-flight bytes is increased by the size of the packet. If an entry is not found, the switch creates a new entry for that flow and stores it in the table. Then, it selects a new output port (according to the destination), stores it in the Flowcut table, and forwards the packet to this port, thus creating a new \name. 

On fat tree networks, \name switching uses up/down routing and selects, in the \textit{up} direction, the least loaded port. On Dragonfly networks~\cite{dragonfly,278358}, \name switching uses UGAL routing for selecting the least loaded path~\cite{singh2005load,dragonfly}. \Name switching is independent from the specific algorithm used for selecting the output port, and any other adaptive routing algorithm can be used. In particular, for UGAL, the choice whether to route subflows minimally or not is still done by the global adaptive algorithm irrespective of Flowcut, which is orthogonal to it. 
For each \name, the switch keeps track of the number of in-flight bytes that are still being routed in the network between itself and the egress switch. For this reason, after the egress switch forwards the packet to the destination host of the flow, it sends back an acknowledgment packet (\textit{ACK} for short) to the ingress switch. 
Each switch forwards the ACK packet on the input port on which the data packets of that \name were received, thus forwarding the ACK through the same switches crossed by the data packet (but in the reverse order). 

An ACK packet contains the key of the flow it belongs to, the size (in bytes) of the corresponding data packet and, a timestamp, and a counter (discussed later), for a total of 20 bytes (we provide a more detailed analysis in Section~\ref{sec:evaluation:consumption:bandwidth}). Because existing networks have a \textit{Maximum Transmission Unit} (MTU) of a few thousand bytes, ACKs introduce a negligible bandwidth overhead.

When a source switch receives an ACK packet, the in-flight bytes counter for the corresponding \name is decreased by the number of bytes carried by the ACK packet. When the counter drops to zero, the \name entry is removed from the table. By doing so, if a switch receives later a new packet for that flow, it will create a new \name, possibly routing its packets on a different path. By allowing a flow that has no in-flight packets to choose a new route, Flowcut switching guarantees in-order delivery of packets while choosing less congested routes for new arriving packets.

We observe that if we want to preserve the in-order guarantee, the first packets that are sent out before receiving any ACK back must be all sent using the same path (using ECMP for example) in order to avoid any possibility of generating out-of-order packets. If, for example, we initially send out a BDP (Bandwidth-delay product) worth of data, then they need to follow the same path. Only once we start getting ACKs back the normal Flowcut switching mechanism starts to kick in.


\begin{figure}[htpb]
\centering
\includegraphics[width=\columnwidth]{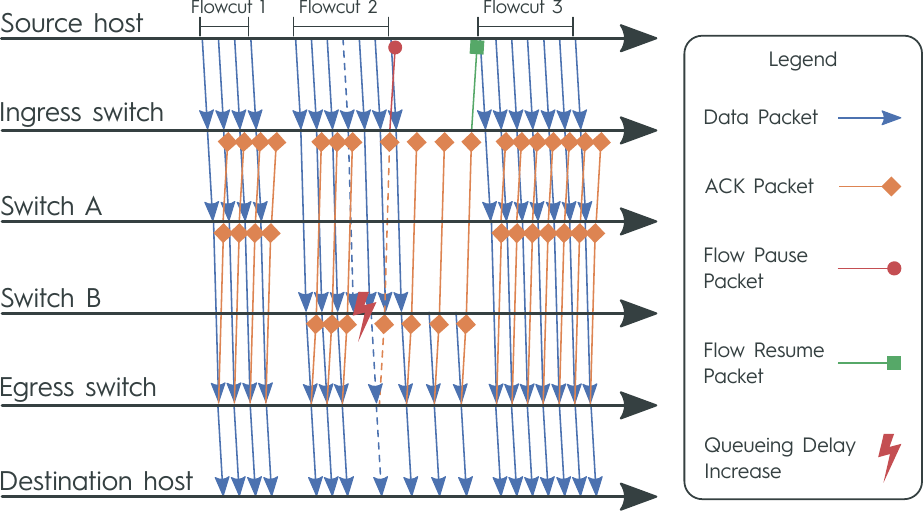}
\caption{\Name switching example.}
\label{fig:flowcut_diagram}
\end{figure}

To better describe how \name switching works, let us consider the example in Figure~\ref{fig:flowcut_diagram}, showing the packets exchanged between the host and the switches depicted in Figure~\ref{fig:flowcut_topology}. The source host sends the first packet (data packets are denoted with \pktdata), and the ingress switch decides to forward it to \textit{Switch A}. Because the ingress switch receives the three subsequent packets while there are still in-flight packets, it forwards them to \textit{Switch A} again. In the meanwhile, the ingress switch receives four ACK packets (\pktack). When the fourth ACK is received, there are no in-flight bytes for that flow, and the \name entry is removed from the table. When the fifth packet arrives, the ingress switch does not find any entry for that \name in its table, and a new \name is created (\Name 2). This time, the ingress switch decides to send the packet to \textit{Switch B} (e.g., because the output port towards \textit{Switch A} became congested in the meanwhile).

\subsection{Flow Draining}\label{sec:flowchip:draining}
Whereas \name switching guarantees in-order delivery, a flow could always have some in-flight bytes and thus, even if the selected path is congested, switches may not be able to reroute that flow on a different path. This might be an issue for protocols like RoCE that exhibit a more steady behaviour with not too many bursts~\cite{8895760}. To balance the traffic even in presence of steady traffic, if the ingress switch detects that a flow is experiencing congestion, it can decide to force the creation of a new \name, that will potentially be routed on a different path. To do so, the switch asks the source of the flow to stop sending packets for that flow. This causes the flow to be completely drained (i.e., all the in-flight packets are received by the destination and the ingress switch receives all the corresponding ACKs). When the ingress switch receives the last ACK for that flow (i.e., there are no in-flight packets), it deletes the \name from the table and asks the source of the flow to resume the transmission of that flow. When a new packet of that flow is received, the ingress switch does not find a \name for that flow in the table, and can thus create a new one, to be forwarded on a different (and less congested) path. 

The draining process is determined by two main aspects: how the ingress switch communicates to the source of the flow to pause/resume the transmission of the packets of that flow (that depends on the specific network) and how it decides that a flow needs to be drained and rerouted. Regarding the first aspect, an effective mechanism, called Fine Grain Flow Control (FGFC) \cite{patent_hpe}, is available on the Slingshot~\cite{slingshot} network. With such mechanism the switch returns FGFC frames for flows causing congestion (similarly to pause frames). These identify a flow using either the normal 5-tuple hash or an identifier provided by the NIC. The possible operations are XOFF, XOFF with credit, and XON. An XOFF with credit instructs the NIC that a flow should send a certain amount of data in the pause period. If the NIC ignores FGFC the traffic class will be paused. FGFC is designed to interoperate with an Ethernet NIC managing large numbers of flows, many of which can make progress while others are blocked or being paced. Otherwise, if such a mechanism is not available and cannot be implemented, the ingress switch can rely on \textit{Priority-based Flow Control} (PFC) packets~\cite{pfc}, thus pausing all the flows belonging to the same traffic class of the flow to be drained.

To estimate the congestion, the ingress switch adds a timestamp to the packet before forwarding it (e.g., by inserting a custom header between Ethernet and IP headers). Then, the egress switch removes this additional header from the data packet before forwarding it to the destination host, and copies the timestamp in the ACK packet. We note that this process is done completely in hardware by the switch with no cpu utilization or increase in the end-to-end latency \cite{slingshot, nvidia_latency}.
Moreover, to address the extra space in the header due to the timestamp, the advertised MTU size is slightly reduced. This is similar to the approach taken by Swift where 4 bytes are used by timestamp in the packet header \cite{swift}.
Every time the ingress switch receives an ACK packet, it computes the round trip time (RTT) $\tilde{r}$ by computing the difference between the current timestamp and the one carried in the ACK packet. Because packets can have different size, the switch could observe an increase in $\tilde{r}$ due to larger transmission latency for larger packets, even without any increase in the queueing latency. For this reason, the switch removes from $\tilde{r}$ the transmission latency, by subtracting the quantity $p \cdot h \cdot t$, where $p$ is the packet size, $h$ is the number of traversed hops, and $t$ is the transmission rate of the switch. To compute $h$ each data packet also carries a counter (incremented at each hop) representing the number of traversed hops, and copied back in the ACK packet by the egress switch. 
We note that, with low diameter topologies where multiple paths are present between the sender and the receiver, it is important that the ACK takes the same reversed path as the corresponding data packet. To do so, we propose two possible solutions: the most straightforward one is to add a single extra field in the switch entry to indicate the input port from which the packets of the flow have been received. That port will be used as the next hop when the ACK is received, thus allowing ACKs to follow the reverse path. The other approach is to move the entire implementation to the NIC (more details in Section \ref{sec:variant:nic}) since the intermediate paths then become of no importance for Flowcut.

To assess if a packet experienced a too high queuing latency, the switch needs to compare the measured RTT to some reference RTT. For this reason, the switch keeps track of the minimum observed RTT $r_{min}(h)$ for each hop count $h$. It is worth remarking that this information is globally stored and does not need to be stored for each flow. Because existing data center networks have a diameter lower than ten~\cite{6552917,slingshot}, this process requires storing only a few extra bytes in the switch memory. 

\begin{table}[htbp]
    \footnotesize
    \small
    \begin{tabular}{c l} 
     \textbf{Param.} & \textbf{Description} \\ 
     \toprule
     {$l$}          & \makecell[l]{Average packet latency (ingress to egress switch).} \\ 
     {$b$}          & \makecell[l]{Average rate of data packets.} \\ 
     {$a$}          & \makecell[l]{Average rate of ACK packets.} \\ 
     {$r$}          & \makecell[l]{Average RTT.} \\ 
     {$\tilde{r}$}  & \makecell[l]{Measured RTT for a packet.} \\ 
     {$r_{min}(h)$} & \makecell[l]{Min. observed RTT for packets crossing $h$ hops.} \\ 
     {$h$}          & \makecell[l]{Number of hops traversed by a packet.} \\ \bottomrule
    \end{tabular}
      \vspace{2pt} 
    \caption{Description of used parameters.}
    \label{tab:model}
\end{table}

Then, the switch can compute a normalized RTT (always $\geq 1$) as $\frac{\tilde{r}}{r_{min}(h) + p\cdot h \cdot t}$. To avoid reacting to temporary increases in the RTT due to transient congestion, the switch stores, for each flow, an exponential moving average $r$ of the normalized RTTs. If the average exceeds a threshold, the flow is drained and rerouted. Limiting the RTT also limits the number of in-flight packets, thus in turn limiting the maximum time required to drain a flow. To proactively react to congestion, the switch also keeps a moving average of the difference between the current and the previously measured normalized RTT. If the average exceeds a threshold, the congestion on the path is growing too quickly and the switch drains and reroutes the flow currently being forwarded. We summarize the parameters used in our discussion in Table~\ref{tab:model}.

It is worth remarking that although the measured RTT also includes the queuing latency experienced on the reverse path by the ACK packets, this is usually negligible because ACK packets can be forwarded with higher priority with respect to data traffic, as it happens in the Slingshot~\cite{slingshot} network. Moreover it has been shown that, in the presence of congestion on the reverse path, RTT measurements with ACK prioritization are indistinguishable from measurements taken in a scenario with no congestion on the reverse path~\cite{timely}. 

Some existing congestion control algorithms also use the RTT as a congestion signal~\cite{timely,swift}. However, while in those cases this is used to reduce the transmission rate of the sender so that it matches the bandwidth of the path, in our case we use it to trigger the selection of a better path, hopefully with a higher bandwidth. We elaborate more on the interaction between the adaptive routing and congestion control algorithms in Section~\ref{sec:design:discussion:cc}.

To better describe the draining process, we report an example in Figure~\ref{fig:flowcut_diagram}. When the fourth packet of \textit{\Name 2} (denoted with a dashed arrow) arrives at \textit{Switch B}, it experiences some delay (\congestion), e.g., because packets of other flows are already queued for transmission to the egress switch. As a consequence, the transmission of the corresponding ACK is delayed as well. When the ingress switch receives the ACK, it realizes that the packet experienced congestion (because its normalized RTT exceeded the threshold).  

Thus, it asks the source of the flow to not send any other packet for that flow (\pktpause). Eventually, when the ingress switch receives all the ACKs for the in-flight packets, it removes the \name from the table and resumes the packets transmission (\pktresume). The source starts transmitting the packets again and, because there are no entries for that flow in the table of the ingress switch, the switch creates a new \name, routing it through a less congested switch (\textit{Switch A}).

Finally, we observe that the draining logic generalizes from \textit{ingress only} to a \textit{full switch} implementation that accounts for per-hop latency. While it increases memory requirements and the number of timestamping operations, it achieves the highest precision and effectiveness.

\subsection{Switch Driven Variants}\label{sec:variant:switches}
While so far we described a full switch implementation of \name{}, the design also supports an ingress-switch only implementation that is attractive for incremental rollouts and for fabrics with tight memory budgets.

\begin{description}[style=unboxed,leftmargin=0pt,labelindent=0pt,labelsep=0.5em]
  \item[\textbf{Full switch}] Every switch along the path maintains per flow state and may reroute locally. This yields the fastest reaction and highest precision at the cost of memory across the fabric.

  \item[\textbf{Ingress only}] Only the ingress switch keeps per flow state and decides when to reroute. Core and aggregation switches forward packets without \name{} state. Because intermediate switches do not update counters for packets in flight, ACK packets do not need to follow the exact reverse path; they can return along any path from the egress to the ingress by carrying and the ingress progresses ACKs before triggering rerouting. In practice, each top of rack switch holds state only for flows sourced by its directly attached hosts.
\end{description}

\subsection{NIC-only Implementation}\label{sec:variant:nic}

\begin{table}[t] 
  \centering

  \renewcommand{\arraystretch}{1.15}
  \begin{tabularx}{\linewidth}{@{} l
                                >{\raggedright\arraybackslash}X
                                >{\raggedright\arraybackslash}X
                                >{\raggedright\arraybackslash}X
                                >{\raggedright\arraybackslash}X @{}}
    \toprule
    \textbf{Variant} &
    \textbf{Memory impact} &
    \textbf{Adaptivity} &
    \textbf{ACK routing} &
    \textbf{Deployment effort} \\
    \midrule
    Full switch &
    Per-flow entry in every switch (highest memory requirements) &
    Re-routing capabilities at all switches for best performance &
    Must follow reverse path storing the input port at switches &
    Upgrade all network switches \\

    \addlinespace[0.4em]
    Ingress only &
    State only in ingress switches (medium memory requirements) &
    Reroute only at ingress, slower and less precise &
    Free to take any path &
    Upgrade ToR; core untouched \\

    \addlinespace[0.4em]
    NIC only &
    No switch state; per-flow info lives in endpoints (lowest memory requirements) &
    Adaptivity limited to ECMP hash change at source &
    Free to take any path &
    Flowcut-aware NIC firmware/driver; fabric unchanged \\
    \bottomrule
    
  \end{tabularx}
  \vspace{1pt} 
  \caption{Qualitative comparison of Flowcut deployment variants}
  \label{tab:flowcut_variants}
\end{table}

In the previous section, we assumed \name switching to be implemented by the switch, without any support from the NIC (except for PFC-like mechanisms). Due to its simplicity, it could be implemented on most existing programmable switches~\cite{p4-survey-1,p4-survey-2}. However, \name switching could also be entirely implemented in hosts's NICs (e.g., by using programmable NICs~\cite{pspin,bluefield,pensando,liquidio,stingray,fungible}) without any switch support, similarly to \textit{Flowbender}~\cite{10.1145/2674005.2674985} or \textit{REPS}~\cite{bonato2025repsrecycledentropypacket}. The switch would not need to store any information about the active flowcuts, and the NICs would only need to store information about the \names it generates, thus also reducing resource occupancy. Because NICs usually have larger memory than switches (and could also use the host DRAM~\cite{pspin}), this would allow the use of \names on networks with larger RTTs. 

For example, the source NIC would add to the packet the timestamp, that would be copied back by the destination NIC in an ACK packet. The source NIC then would compute the RTT and, if the current path is congested, it would stop the transmission of packets for that flow until all the ACKs are received. If the switches use ECMP, the source NIC could force the selection of a different path by changing one of the fields on which the ECMP hash is computed (e.g., the source port), so that the packet (and the subsequent ones), will be forwarded on a different path (similar to what is done in SRD~\cite{9167399}). The original source port can be inserted in an additional header, and then restored by the destination NIC. However, in this case, the selection of the path is random and congestion oblivious. Moreover, because there is no support from the switch, ACKs must also have Ethernet and IP headers, thus increasing their network bandwidth occupancy (even if still negligible for MTU-sized data packets).

This flexibility in moving functionalities from NICs to switches and vice versa allows incremental deployment of \name switching, either by implementing it on programmable NICs without changes in the switches, or by implementing it in top-of-the-rack (ToR) switches without any change at the endpoints.

In Table~\ref{tab:flowcut_variants} we summarize the different variants of Flowcut with their different advantages and disadvantages.

\section{Evaluation}\label{sec:evaluation}
In this section we first analyze the resource requirements of \name switching in terms of network bandwidth and switch memory (Section~\ref{sec:evaluation:consumption}). We then describe our simulation environment (Section~\ref{sec:evaluation:setup}) and evaluate \name switching on different networks and workloads, by comparing it with other state of the art adaptive routing algorithms (Section~\ref{sec:evaluation:simulations}). Finally, we share some results based on a real cluster running Slingshot on 2,048 endpoints (Section~\ref{sec:real_hw_run}). 

For Flowcut, we use the \textit{NIC only} version (Section~\ref{sec:variant:nic}) for all experiments unless specified differently. This is to showcase the performance of the easiest to implement version of Flowcut. Finally, in Section~\ref{flowcut_variants_result} we showcase how implementing Flowcut on switches can further improve the performance.

\subsection{Network Resources Occupancy}\label{sec:evaluation:consumption}
\Name switching increases the consumption of network resources, both in terms of network bandwidth due to the transmission of additional information in data and ACK packets, and of switch memory.

\subsubsection{Network Bandwidth}\label{sec:evaluation:consumption:bandwidth}
\Name switching requires explicit ACK packets to be sent from the egress to the ingress switch. To reduce the Ethernet header overhead, ACKs can be sent using a custom packet format, using the first byte of the Ethernet preamble to distinguish them from data packets. ACKs do not have the IP header, because they are routed on the backward path using information stored in the switches (input and output port for the flow). Thus, ACKs are only composed of a preamble byte and a payload containing the key of the flow (13 bytes if we consider the 5-tuple composed of the source and destination IP addresses, source and destination ports, and transport protocol identifier). To further reduce the number of extra transferred bytes, the flow key could be mapped to a smaller space. For example, by transmitting only the least significant bits of the addresses (depending on the subnet mask) or by using a unique flow key generated by the source NIC and encoded on fewer bytes. To be as general as possible, however, we consider the case where all the 13 bytes of the 5-tuple key are transmitted in the ACKs.

\Name switching also needs to add the timestamp and the hop count to both data and ACK packets. Because end-to-end latencies in existing data center networks are in the order of a few microseconds~\cite{swift,slingshot,bfc,278358}, if timestamps have nanosecond resolutions, RTTs can be encoded using 2 bytes (for RTTs up to 64 microseconds). Data center networks have a diameter lower than ten~\cite{6552917,slingshot}, and storing the number of hops only introduces a 4 bits overhead in both the data and ACK packets. To align the packet size to 1 byte, additional 4 bits are unused and reserved for future extensions. Alternatively, on IP networks switches could use the \textit{Time to Live} (TTL) field of the IP header instead of adding an explicit counter to the data packet. We summarize the extra network bandwidth consumption in Table~\ref{tab:comparison}. 

\begin{table}[h!]
\centering

\footnotesize
\begin{tabular}{c c c c} 
 {\textbf{Algorithm}} & \makecell[c]{\textbf{Congestion}\\\textbf{Signal}} & \makecell[c]{\textbf{Per-Packet}\\\textbf{Extra Bytes}\\\textbf{(Wire)}} & \makecell[c]{\textbf{Per-Flow}\\\textbf{Extra Bytes}\\\textbf{(Memory)}} \\ 
 \hline
 Flowcell~\cite{flowcell}       & None      & 0 & 2  \\
 Flowlet~\cite{letitflow,conga} & Idle Time & 0 & 5  \\
 \Name Full switch  & RTT           & 20 & 11
 \\
 \Name Ingress/NIC only  & RTT           & 20 & 10 \\
 \hline
\end{tabular}
\vspace{1pt} 
\caption{Resource occupancy of different adaptive routing algorithms. We report the extra bytes sent for each packet, and the extra bytes stored for each flow in the switch memory. We note that although the per-packet or per-flow cost is almost the same for all Flowcut variants, in practice \textit{NIC/Ingress only} require significantly less resources on the NIC/Switch due to less active flows.}
\label{tab:comparison}
\end{table}

For 1KiB packets, the per-packet \name switching network bandwidth overhead is smaller than 2\%. We note that modern switches, such as the Rosetta switches used in Slingshot \cite{slingshot}, are capable of supporting per-packet ACKs even at modern network bandwidths ($\ge 200 Gb/s$).

In principle, for protocols like TCP we could avoid sending explicit ACKs and piggyback the additional information on top of the TCP ACKs. This has, however, several disadvantages. For example, relying on TCP ACKs would delay the reception of the congestion feedback on the switch, due to the extra latency required to cross the receiver host network stack. This latency would increase even more if the TCP ACK also carries data (due to the additional transmission latency at each hop) or if ACK coalescing is performed by the receiver host. Having said that, Flowcut would still be able to work, although slightly delayed, even in cases where per-packet acking is not feasible and cumulative ACKs are used.

\subsubsection{Switch/NIC Memory}\label{sec:evaluation:consumption:memory}
\Name switching, when implemented in all switches, require switches to store, for each flow crossing them, the input and output ports for the flow (1 byte each for 256-port switches) and the number of in-flight bytes (3 bytes for up to 2 MiB of in-flight data per flow), for a total of 5 bytes per flow. The switch also needs to store the normalized RTT moving average, that requires 2 bytes, as discussed in Section~\ref{sec:evaluation:consumption:bandwidth}, the normalized RTT measured for the last received ACK, and an exponential average of the RTT difference, requiring an additional 4 bytes per flow. For the Ingress only and NIC-based implementations, Flowcut does not need to store the input port saving one byte. This is because in most topologies the NICs are linked only to one switch and the ACK packet will necessarily travel to that switch regardless of what path it takes to get there. We summarize the memory requirements in Table~\ref{tab:comparison}. We also report the per flow state for Flowlet switching, that requires 5 bytes per flow~\cite{letitflow,conga}, and for flowcell switching, that requires 2 bytes per flow~\cite{flowcell}.

The memory occupancy depends on the number of active flows (i.e., flows that have at least one in-flight packet). Indeed, the NIC or a switch deletes the information about a \name from the table as soon as there are no in-flight packets for that flow. To simplify the analysis, we assume that all the transmitted packets have size $M$ (equal to the MTU). If each host has $f$ flows, and the network bandwidth $B$ is equally divided among these flows, each of the flow has a bandwidth $b = \frac{B}{f}$. The number of in-flight packets for a flow can be computed as the bandwidth-delay product. Thus, if the packet latency is $l$, each flow has $\frac{b \cdot l}{M}$ in-flight packets. 

If each flow has at least one in-flight packet, all the $f$ flows are active. Otherwise all the flows have, on average, less than one in-flight packet. Each host can have at most $\frac{B \cdot l}{M}$ in-flight packets, and in this case each in-flight packet belongs to a different flow. Thus, each host would have $\frac{B \cdot l}{M}$ active flows. To summarize, if there are $H$ hosts in the system, there will be a total number of active flows $F$ in the network equal at most to:


\begin{equation}\label{eq:active-flows}
F = \begin{cases} H \cdot f, & \mbox{if } \frac{B \cdot l}{f \cdot M} \geq 1 \\ \frac{H \cdot B \cdot l}{M}, & \mbox{if } \frac{B \cdot l}{f\cdot M}  < 1 \end{cases}
\end{equation}

It is worth remarking that when each flow has, on average, less than one in-flight packet, increasing the number of active flows per host $f$ does not increase the number of active flows in the network. Thus, the maximum number of active flows that can be in the network at any time is only determined by the number of hosts $H$, the network bandwidth $B$, the latency $l$, and the MTU $M$ (assuming that each packet is as large as the MTU).

\begin{figure*}
    \centering
    \includegraphics[width=\linewidth]{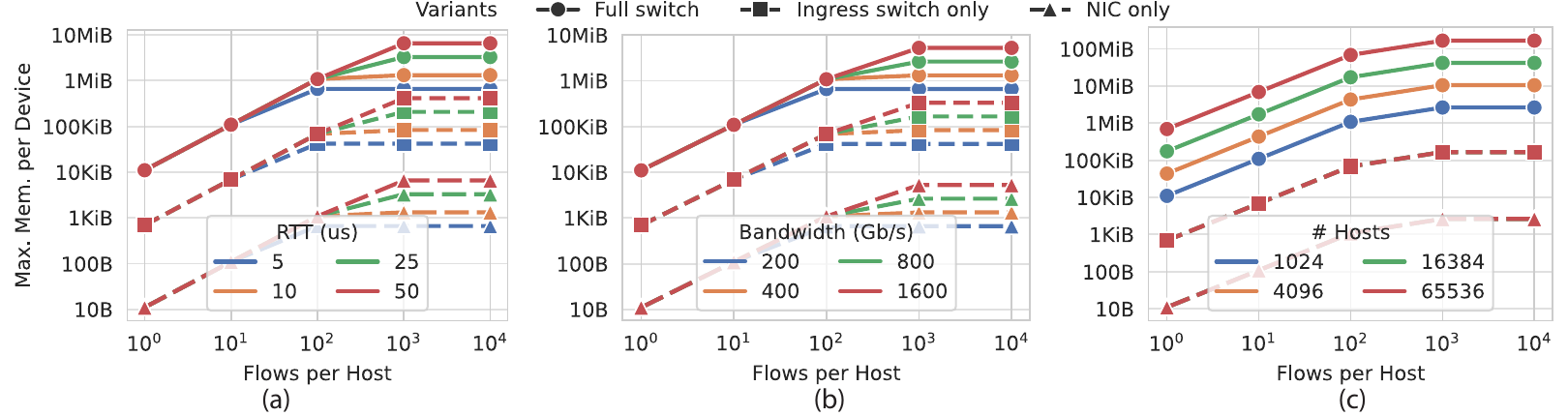}
  \caption{Maximum switch memory occupancy (MiB) of \name switching on different configurations on a network with 2KiB MTU and 64 input and output ports per switch. (a) 1024 hosts on a 200Gb/s network, for different RTTs. (b) 1024 hosts, 5 microseconds maximum RTT, for different network bandwidths. (c) 800 Gb/s network, 5 microseconds maximum RTT, for different hosts count.}
\label{fig:flowchip_memory:mpr}
\end{figure*}

This can also be observed in Figure~\ref{fig:flowchip_memory:mpr}(a), where we report the modelled maximum memory occupancy of \name switching for different values of RTTs, network bandwidth, and number of hosts. The memory occupancy is computed as the total number of active flows (derived from Equation~\ref{eq:active-flows}) multiplied by the number of bytes that a switch needs to store for each flow (Table~\ref{tab:comparison}). For the \textit{Full switch} variant of Flowcut, we consider a worst case scenario where a switch stores information for all the flows in the network (that is usually not the case since each flow crosses only a portion of the switches). On the other hand, the Ingress only variant has a much smaller memory requirement as only flows coming from its input ports need to be saved. Similarly, the \textit{NIC only} version of Flowcut has a much smaller memory requirement since it only needs to store flows active on a given NIC.

In Figure \ref{fig:flowchip_memory:mpr}(a), we fix the number of hosts to 1024, and the network bandwidth to 200 Gb/s, and we analyze how the memory occupancy changes for different RTTs. We can see that memory occupancy grows linearly with the RTT but, for a fixed RTT, when increasing the number of flows per host after a certain point the switch memory occupancy remains constant. Because \textit{North-South} traffic in data centers (i.e., traffic directed to hosts outside the data center) can be characterized by RTT in the order of milliseconds, \name switching is only used for \textit{East-West} traffic (i.e., to traffic between hosts inside the data center). We also observe that even for a latency of 50 microseconds, much higher than that observed for intra-datacenter traffic~\cite{swift,slingshot,bfc,278358}, the memory occupancy on the switches does not grow beyond 7 MiB for the \textit{full switch} implementation. 

In Figure~\ref{fig:flowchip_memory:mpr}(b), we instead fix the number of hosts to 1024, and the RTT to 5 microseconds, and we analyze the memory occupancy for different values of the network bandwidth, ranging from 200 Gb/s to 1.6 Tb/s. Even in this case, we can see that \name switching memory occupancy is limited even for future 1.6 Tb/s networks. Similarly, we analyze in Figure~\ref{fig:flowchip_memory:mpr}(c) the memory occupancy for a 800 Gb/s network with 5 microseconds RTT, for different number of hosts, ranging from 1024 to 65536. In this case we can see that with more than 16384 hosts, the memory occupancy exceeds 50 MiB for the Full switch implementation. However, it is worth remarking that on such a high node count, the \textit{ingress switch only} and the \textit{NIC only} variants of \name switching could be a better solution as they only require slightly more than 100KiB and 10KiB respectively.

\begin{figure}

    \centering
    \includegraphics[width=\columnwidth]{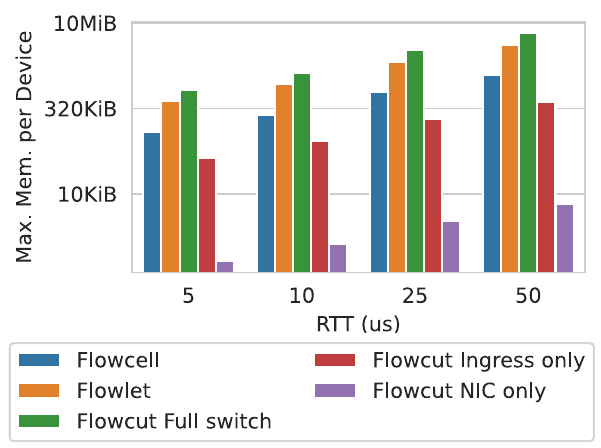}
    \caption{Maximum memory occupancy (MiB) of different algorithms ($10^4$ flows per host), on a 200Gb/s network with 2KiB MTU and connecting 1024 hosts. The memory requirement is for switches for all cases but the NIC only one where it is per NIC.
    } 
    \label{fig:flowchip_memory:all}
\end{figure}

We also report in Figure~\ref{fig:flowchip_memory:all} the memory occupancy of Flowlet and Flowcell switching. If we consider that existing switches have memories of tens of MiB (excluding the packets buffer)~\cite{10.1145/3387514.3405855,10.1145/3098822.3098824}, and that our analysis considers the worst case (with the highest number of active flows on each switch), we believe that \name switching can effectively be implemented with the current technology. Finally, we note that the flexibility of the different Flowcut variants allows users to choose what best fit their needs with the \textit{NIC only} version requiring significantly less space than Flowcell and Flowlet.

\subsection{Test Environment}\label{sec:evaluation:setup}

\subsubsection{Simulated networks}
To analyze \name performance we simulate different types of traffic using the SST simulator~\cite{sst-0,sst}. Hosts in the simulated systems communicate through an RDMA-like protocol, and the network uses credit-based flow control to guarantee lossless behavior (protocols implementing RDMA are usually deployed on lossless networks~\cite{10.1145/2785956.2787484,10.1145/2934872.2934908}). In our analysis we simulated the following systems, all based on 1024 nodes, 200 Gb/s networks and 1us links latency:
\begin{itemize}[style=unboxed,leftmargin=0cm]
\item \textbf{Fat tree (1:1)} This system connects the servers through a 3-level non-blocking fat tree. There are 16 pods with a total of 128 ToR switches and 128 aggregation switches. The aggregation switches are connected to a total of 64 core switches.
\item \textbf{Fat tree (2:1)} This system also connects the servers through a 3-level fat tree, but with a 2:1 tapering. That means that the ToR switches have less (half) up-links going to the aggregation switches.
\item \textbf{Dragonfly} This system has the same configuration of a real system deployed at the \textit{Swiss National Supercomputing Center}~\cite{alps}, using the Slingshot interconnect~\cite{slingshot}. Nodes are interconnected through a Dragonfly network~\cite{dragonfly}. The network is composed of 64-port switches. Servers are divided into four Dragonfly groups (256 servers per group). Switches within a group are fully connected, and each switch is connected to 16 servers, as it happens in the Slingshot interconnect~\cite{slingshot}. Each group is connected to each other group through 16 links. 
\end{itemize}

\subsubsection{Workloads}
In our experiments we consider the following workloads:
\begin{itemize}[style=unboxed,leftmargin=0cm]
\item \textbf{Web Search/Enterprise/Alibaba/Random distribution} In these workloads, each host iteratively selects a random partner and sends a message to that partner. The size of the messages is extracted from distributions (Figure~\ref{fig:distr}) collected from real data centers, and used to evaluate several adaptive routing algorithms~\cite{letitflow,conga}. On fat tree networks, we adopt the same approach used by CONGA~\cite{conga}, forcing all the traffic to cross aggregation switches.
\begin{figure}
    \centering
    \includegraphics[width=\columnwidth]{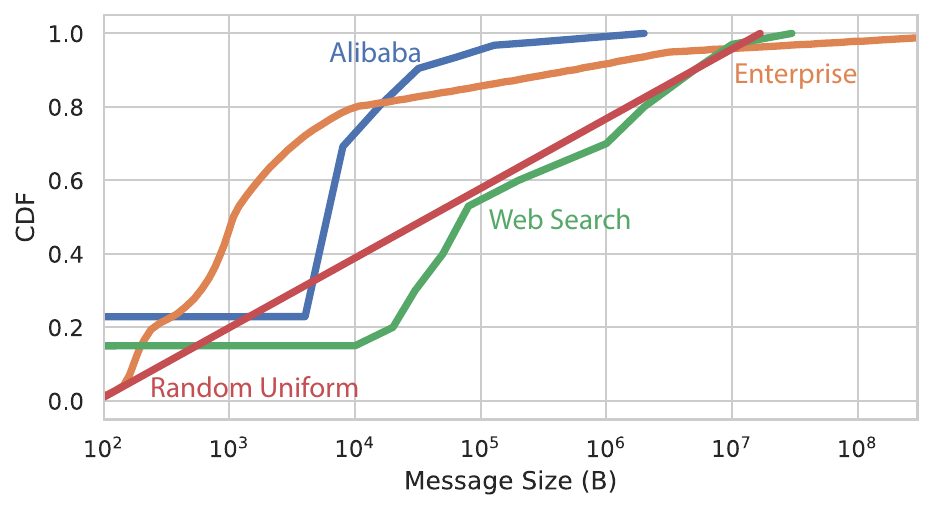}
    \caption{Flow size distribution for some of the different workloads used during benchmarking.}
    \label{fig:distr}
\end{figure}

\item \textbf{Permutation} This microbenchmark consists in all nodes of the network sending a fixed amount of data to another random node in the network. Each node interacts with at most 1 node, meaning at any given time there will be one send and one receive active. This pattern is extremely important in many applications and even AI workloads. For example, the butterfly all-reduce operation uses several permutation operations during its execution. The same can be true for some all-to-all implementations, for instance when using a windowed algorithm \cite{butterfly}.

\item \textbf{All-to-all} An All-to-all benchmark to stress the network. Note that all-to-all collectives are used in a large number of applications including as a training step for the latest deep learning applications \cite{alltoall, ml_alltoall}. Moreover, all-to-all collectives are also frequently used in high-performance computing (HPC) \cite{alltoallhpc}.

\end{itemize}

\subsubsection{Failures}
One important aspect of adaptive routing is correctly handling failures. Practically this means avoiding the utilization of failed links as much as possible. We simulate network failures by downgrading the capacity of a number of links (we show results for 1\% of link failures) to a bandwidth of 1/10th of the initial capacity (20Gbps). We showcase results for some of the experiments and configurations previously mentioned.

\subsection{Experimental Results}\label{sec:evaluation:simulations}
\subsubsection{Tuning Flowcut parameters}
\begin{figure*}[ht!]
\centering
\includegraphics[width=.33\textwidth]{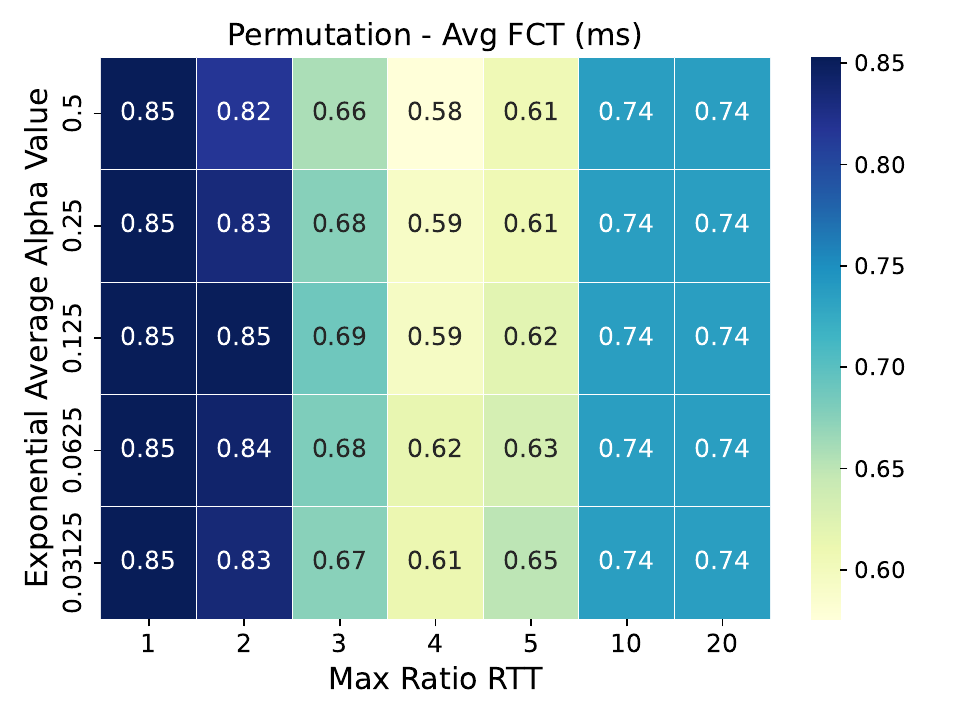}\hfill
\includegraphics[width=.33\textwidth]{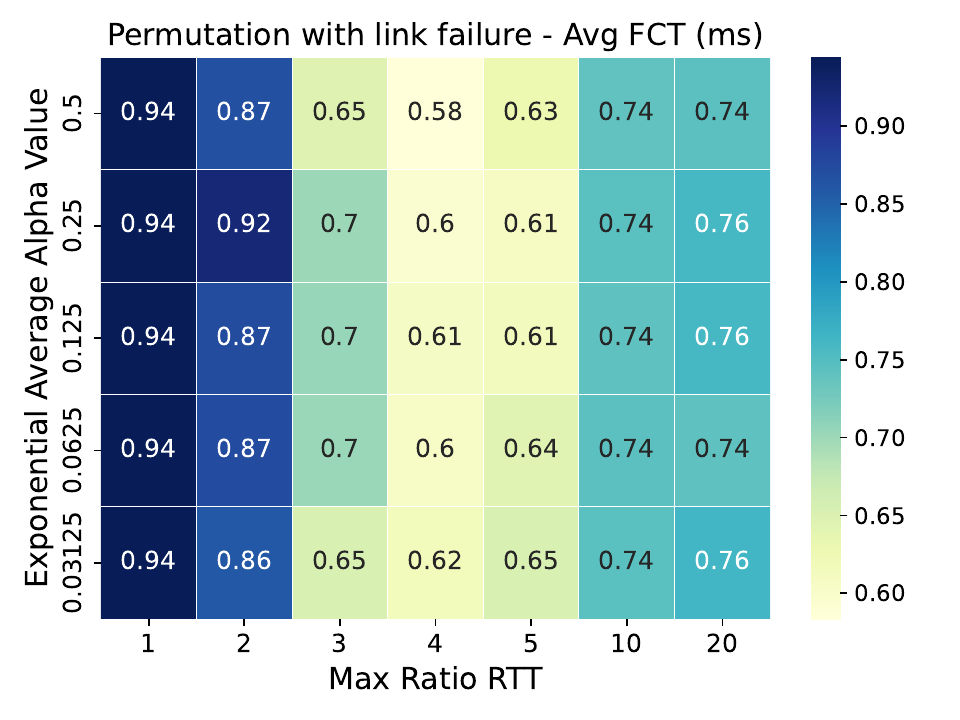}\hfill
\includegraphics[width=.33\textwidth]{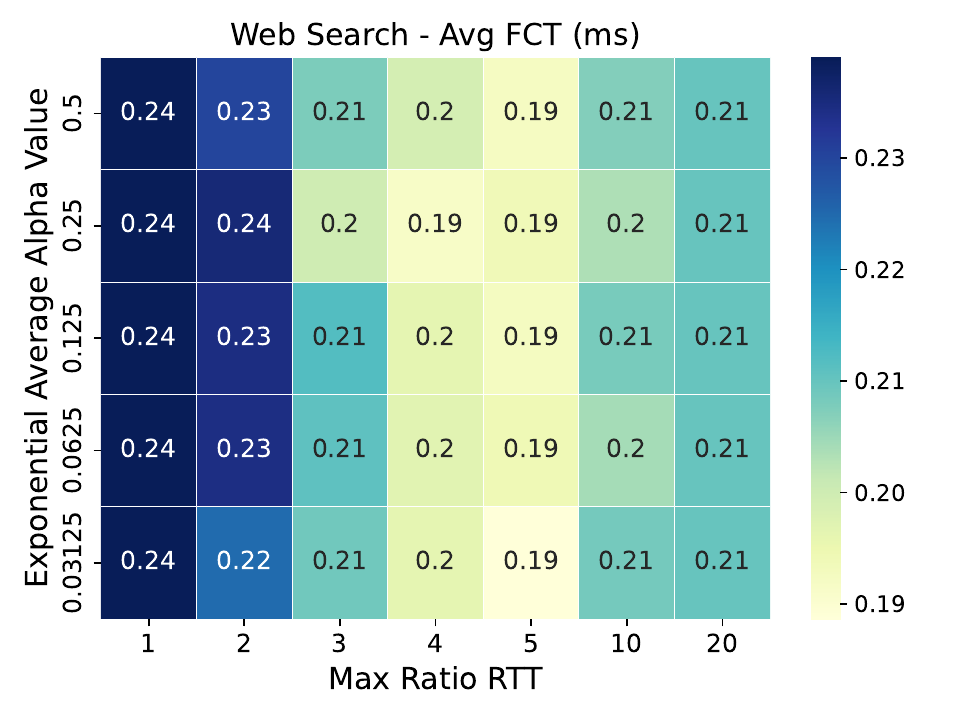}
\caption{Heatmap observing how changing the RTT ratio together with the $\alpha$ value from the exponential average affects runtime.}
\label{fig:heat}
\end{figure*}
As a first evaluation step we run simulations to observe the behaviour of Flowcut as we change some of its most important parameters previously introduced in  Section~\ref{sec:flowchip:draining}. We show in Figure~\ref{fig:heat} the results using a different set of workload and network conditions. We do this to demonstrate that, regardless of the workload or the network's configuration and conditions, Flowcut switching can work effectively without needing extensive parameter tuning.

For each heatmap, on the x-dimension we show different values for the RTT threshold triggering the draining. 
Because Flowcut computes the exponential moving average of the last observed RTT values as $r = \alpha \tilde{r} + (1-\alpha)r$, on the y-dimension, we show different values of $\alpha$ to analyze Flowcut sensitivity to this parameter. Each heatmap reports the average flow completion time in the cells (darker colors represent higher runtimes).

First, we observe that, for all workloads and conditions, $\alpha$ has a minimal impact compared to the RTT threshold. Still, having large values for $\alpha$ introduces some benefits since \name can react to congestion faster. Regarding the RTT threshold, we briefly recall that Flowcut is triggered when the ratio between the average RTT and the minimum observed RTT exceeds such threshold. We observe that a small value always leads to the worst performance. This is expected, because the algorithm would drain flows too often, hence introducing a high overhead from the draining operation. Specifically, a value of $1$ means that the algorithm drains a flow every time the measured RTT is higher than the minimum RTT. 
A similar conclusion can be made for a value of $2$ where the draining time would still introduce large overhead to the overall runtime and flow completion time. 
On the other hand, when setting a large RTT threshold Flowcut switching would never or rarely react to congestion, thus mimicking ECMP behaviour. 

In general, differently from Flowlet Switching, any value in a reasonable range (e.g., between 3 and 5 for the RTT threshold) always leads to good performance, as we show in the following section when comparing these results to other algorithms. It may be counter-intuitive that choosing a large value for the RTT ratio threshold (like 4 or 5) still produces good results. However, such a choice has the advantage of having a minimal draining overhead, while also dealing with the most severe congestion scenarios in the network. Last, we obtained comparable results also when running the same analysis on different network topologies (with different over-subscription ratios).

\subsubsection{Analysis on fat tree networks} \label{sec:fat_tree_results}
\begin{figure}[ht!]
\centering
\includegraphics[width=.5\columnwidth]{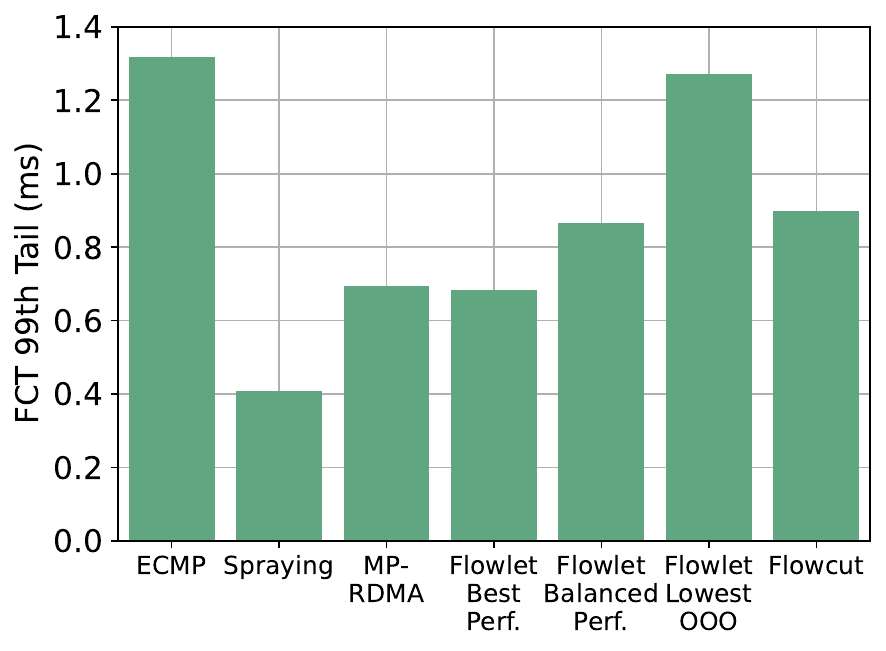}\hfill
\includegraphics[width=.5\columnwidth]{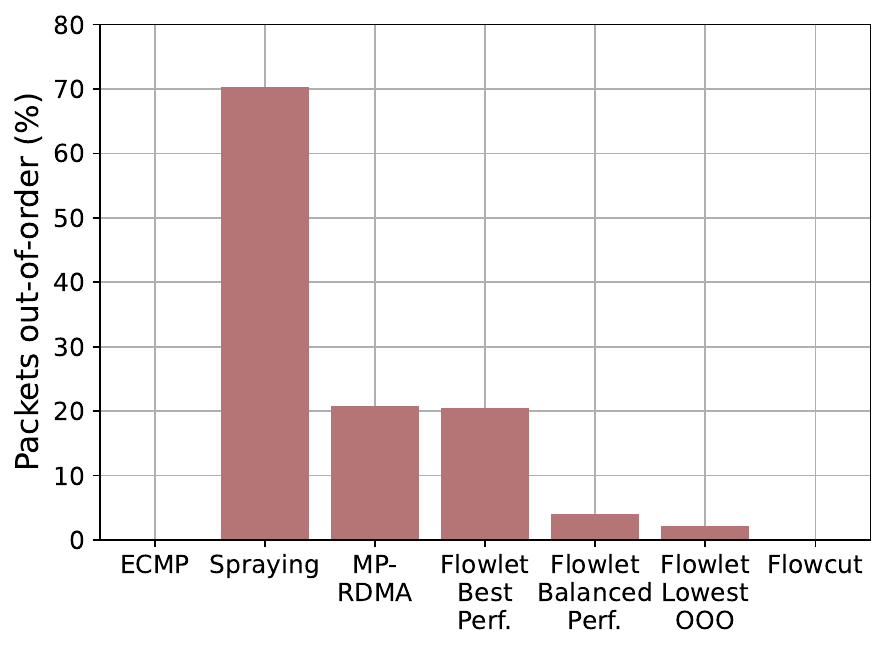}
\caption{99th percentile FCT and amount of out-of-order packets when running a permutation on fattree (untapered).}
\label{fig:ft1}
\end{figure}

We now compare \name with other state of the art algorithms on fat tree networks. 
We define a packet as arriving out-of-order if the expected PSN does not match the received one. Since Flowlet switching performance significantly changes depending on the choice of its parameters, we consider in the analysis the following alternatives. In \textit{"Flowlet Best Performance"}, the parameters have been tuned so that Flowlet switching can provide the best performance while keeping the number of out-of-order packets below 20\%. Likewise, \textit{"Flowlet Balanced Performance"} keeps the number of out-of-order packets below 5\%, and \textit{"Flowlet Lowest OOO"} below 2\%. 
For Flowcut switching, we select an RTT ratio of 4, meaning a flow will be re-routed when the measured RTT is 4 times larger than the base one. \textit{Spraying} is the packet spraying algorithm~\cite{spraying}, distributing each packet randomly across the available paths. This is close to ideals in fat tree networks when there are no failures and each path is equivalent to the alternatives in length. 
MP-RDMA \cite{multipath} is a multi-path RDMA transport that marries ECN-based congestion control with per-packet ACK clocking to continuously steer traffic onto lightly loaded paths and prune underperforming ones, thus bounding out-of-order packet arrivals. A single tunable threshold $\Delta$ lets you cap the maximum tolerated reordering.
We note that we do not consider the impact of re-ordering in this experiments, thus presenting a worst case situation for Flowcut switching as all other algorithms (expect ECMP) would need extra-overhead to deal with the re-ordering of packets.
In the first experiment (Fig~\ref{fig:ft1}), we showcase the results of a 8 MiB permutation using a non-blocking fat tree. We report the 99th percentile average flow completion time and number of out-of-order packets. We focus on the 99th percentile since tightly coupled operations complete only when the last flow has finished. Although packet spraying minimizes the FCT, more than 50\% of packets arrive out-of-order. On the opposite extreme, ECMP has the largest FCT, but no out-of-order packets. Flowcut switching, with its in-order delivery guarantee, is characterized by a FCT lower than \textit{Flowlet Lowest OOO} and on-par with \textit{Flowlet Balanced Performance}. MPRDMA performs slightly better than Flowcut but can't guarantee that all packets are delivered in-order. We note that, in all the experiments, the results for Flowcut also includes the time spent draining (more details in Section~\ref{sec:discussion:drainingtime}), while we assume a best case scenario (worst for Flowcut) of zero re-ordering cost for all the other algorithms.

\begin{figure}[ht!]
\centering
\includegraphics[width=.5\columnwidth]{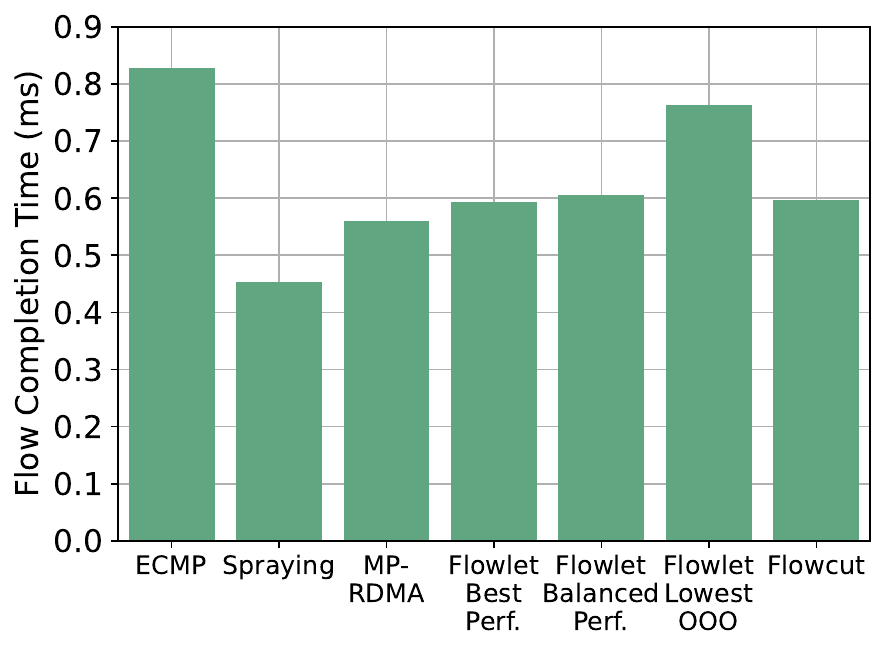}\hfill
\includegraphics[width=.5\columnwidth]{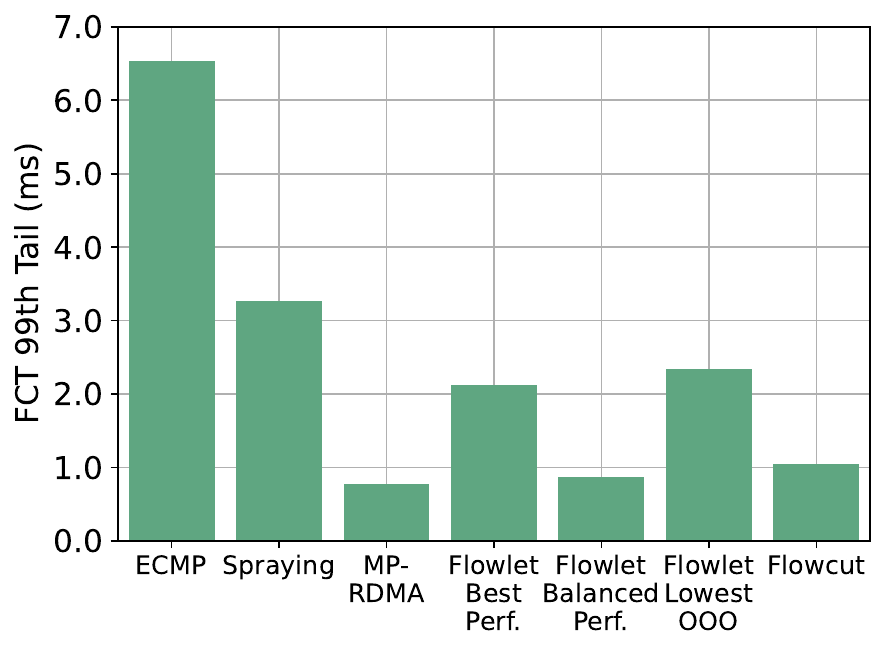}
\caption{Average FCT and 99th percentile FCT when running a permutation on fattree while also simulating few links failing.}
\label{fig:ft_fail}
\end{figure}

We now run the same experiment but by disconnecting 1\% of the links, as previously described. We show in Figure~\ref{fig:ft_fail} the results about the average FCT and the 99th percentile FCT.
We observe how, in this case, the average flow completion time confirms our previous findings with Flowcut getting very similar results to \textit{Flowlet Best Performance}, \textit{Flowlet Balanced Performance}, and MPRDMA. This is expected as the average flow is not affected by the failure for the most part and hence, the results are not much different compared to Figure~\ref{fig:ft1}. However, when looking at the 99th FCT, the situation becomes the opposite with Flowcut and Flowlet competing for the best performance, while Spraying and ECMP lag behind. Indeed, they are both unable to react to link failures. On the other hand, both Flowlet and Flowcut can somewhat successfully circumvent failed links. Flowcut does this by observing the RTT of the packets going through the failed links increasing, while Flowcut achieves this by re-routing packets that arrive with enough delay between them. Interestingly, this also makes the choice of the Flowlet timeout even more challenging, as in this case all values perform very similarly for the tail latency but not for the average latency where they follow the same pattern we will see throughout these results. 

\begin{figure}[ht!]
\centering
\includegraphics[width=.5\columnwidth]{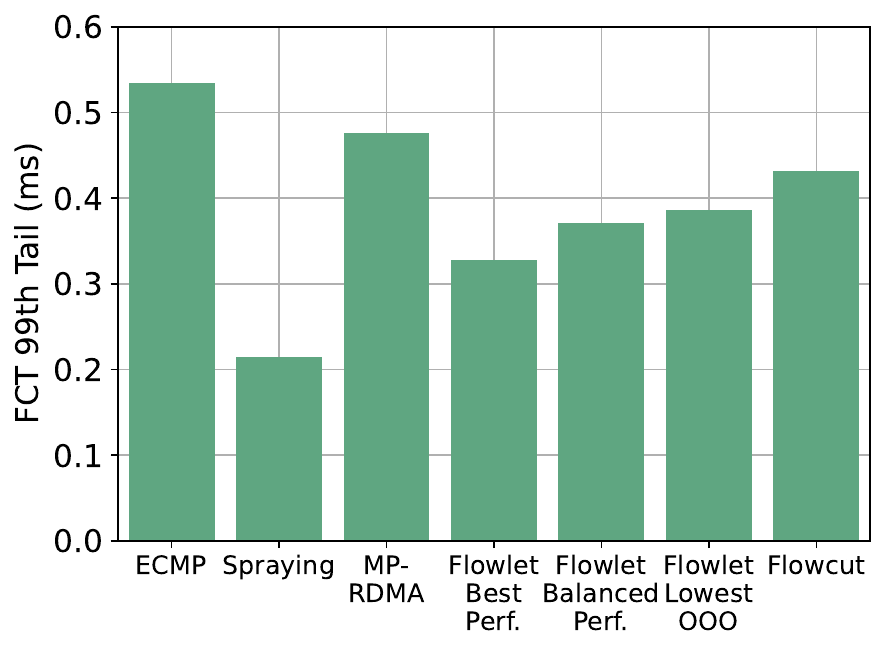}\hfill
\includegraphics[width=.5\columnwidth]{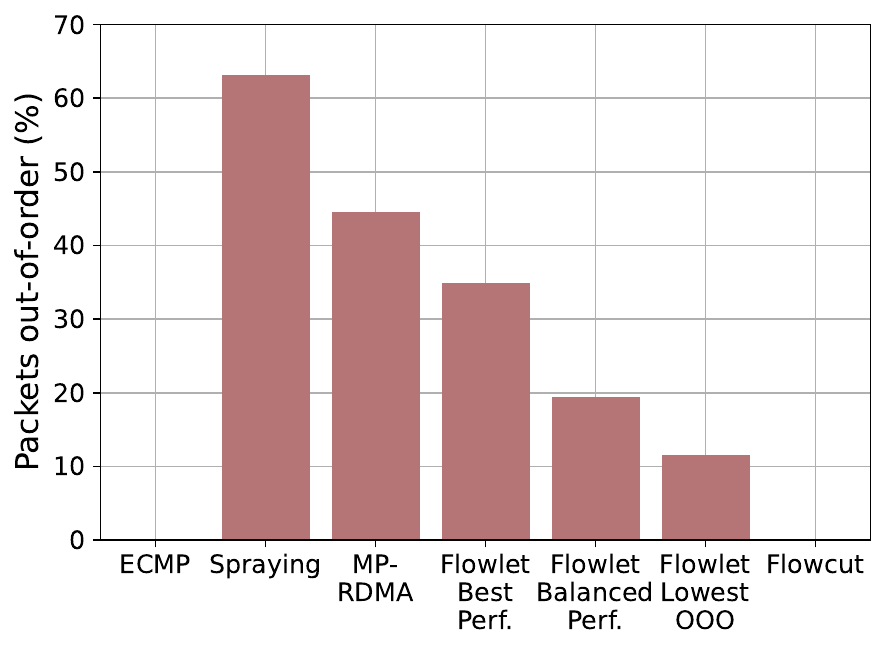}
\caption{99th percentile FCT and amount of out-of-order packets when running an all-to-all on fattree (untapered).}
\label{fig:alltoall}
\end{figure}

We then simulate a 1 MiB all-to-all using the non-oversubscribed topology and without any failure. We show the results in Figure~\ref{fig:alltoall} showing, once more, the 99th percentile flow completion time and the amount of out-of-order packets.
Once more, we confirm the previous findings even during more intensive collectives with Flowcut switching being able to have a tail latency comparable to the one of \textit{Flowlet Balanced Performance}. Interestingly, we note that MPRDMA performs slightly worse in this case, likely due to the different congestion control handling rather than pure load balancing.

\begin{figure}[ht!]
\centering
\includegraphics[width=.5\columnwidth]{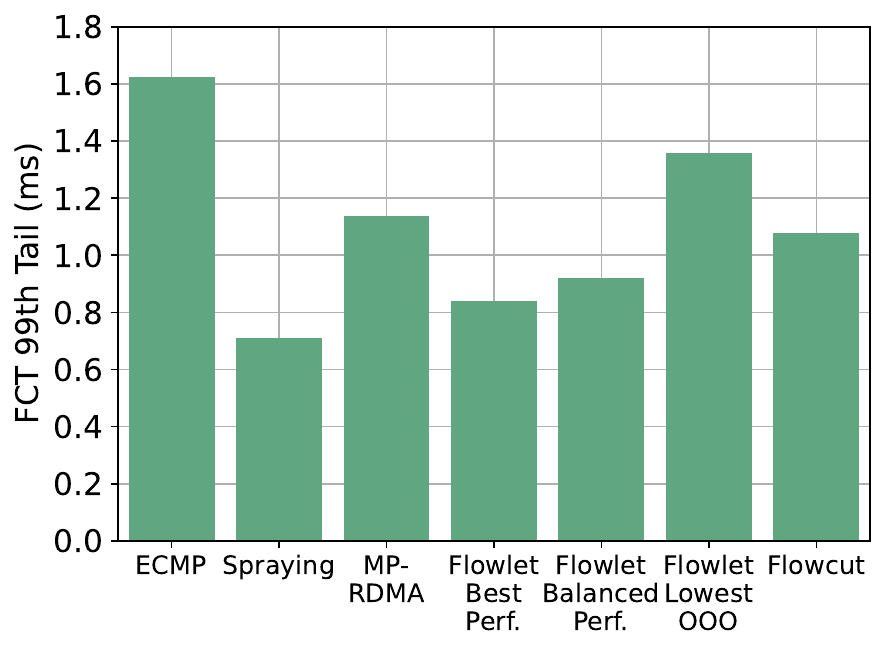}\hfill
\includegraphics[width=.5\columnwidth]{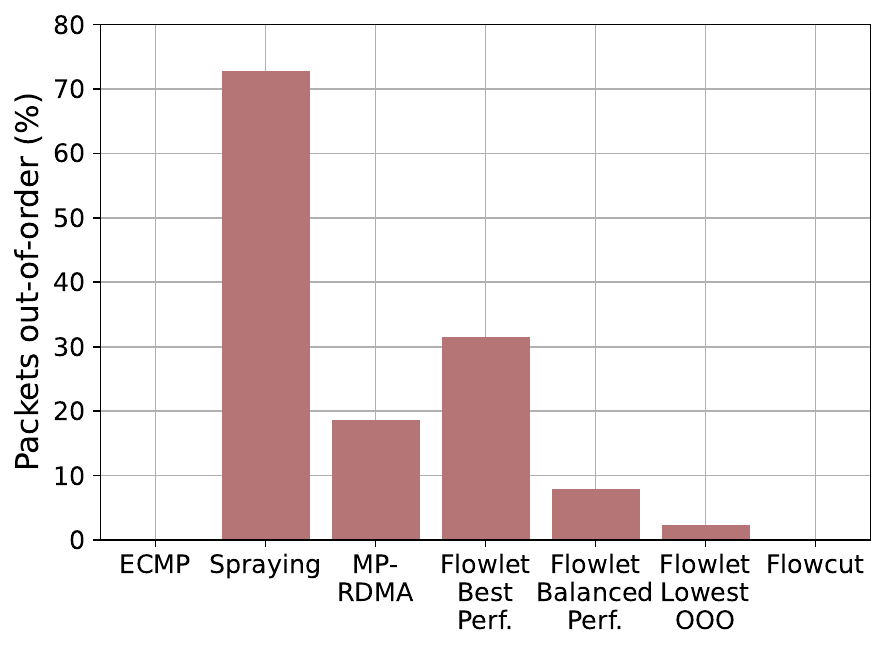}
\caption{99th percentile FCT and amount of out-of-order packets when using a random uniform distribution on a 2:1 oversubscribed fattree.}
\label{fig:ft_os}
\end{figure}

Last, in Figure~\ref{fig:ft_os} we show the results for the oversubscribed fat tree running the data mining distribution, reporting the average flow completion time and the amount of out-of-order packets.

In this case, there is a smaller difference between the routing scheme in terms of FCT. This is because most of the flows, as shown in Figure~\ref{fig:distr}, are smaller and hence, they complete sending their data before the routing algorithm can make any routing adjustment. However, even in this case, we can still see how Flowcut switching outperforms ECMP and all versions of Flowlet switching. Flowcut switching is only slightly outperformed by packet spraying which, however, is characterized by almost 50\% of out-of-order packets. Moreover, in this case MPRDMA performs slightly worse than Flowcut.

\subsubsection{Analysis of Flowcut variants} \label{flowcut_variants_result}

\begin{figure*}
    \centering
    \includegraphics[width=\linewidth]{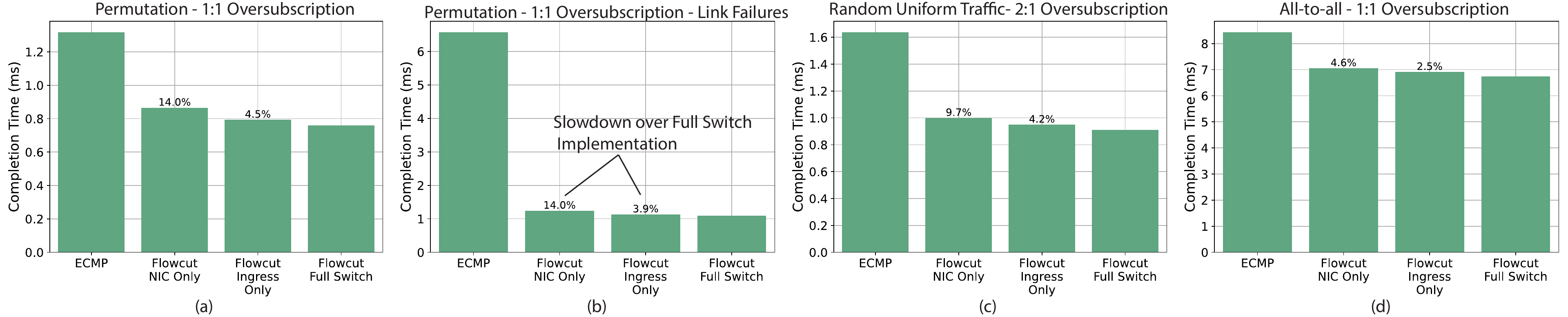}
  \caption{Showcasing the different performance of the three variants of Flowcut. Annotated is the slowdown over the best variant of Flowcut which is always the one implemented on all switches (Full switch).}
\label{fig:variants}
\end{figure*} 

We now focus on analyzing the different variants of Flowcut for the results previously presented for the fat tree topology. In particular in Figure~\ref{fig:variants} we showcase the three variants in 4 different scenarios that we previously presented. We also report again ECMP numbers as a reference. As expected, the Full switch variant is always the best but, interestingly, the NIC version and the Ingress only versions are performing just slightly worse and within 10\% of the full and costly implementation.

\subsubsection{Analysis on Dragonfly}
\begin{figure}[ht!]
\centering
\includegraphics[width=.5\columnwidth]{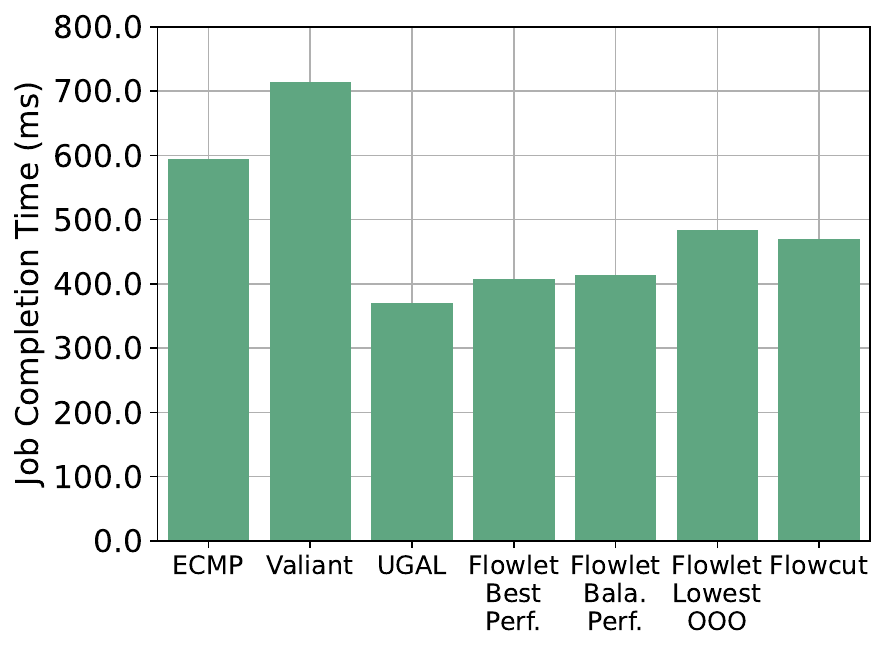}\hfill
\includegraphics[width=.5\columnwidth]{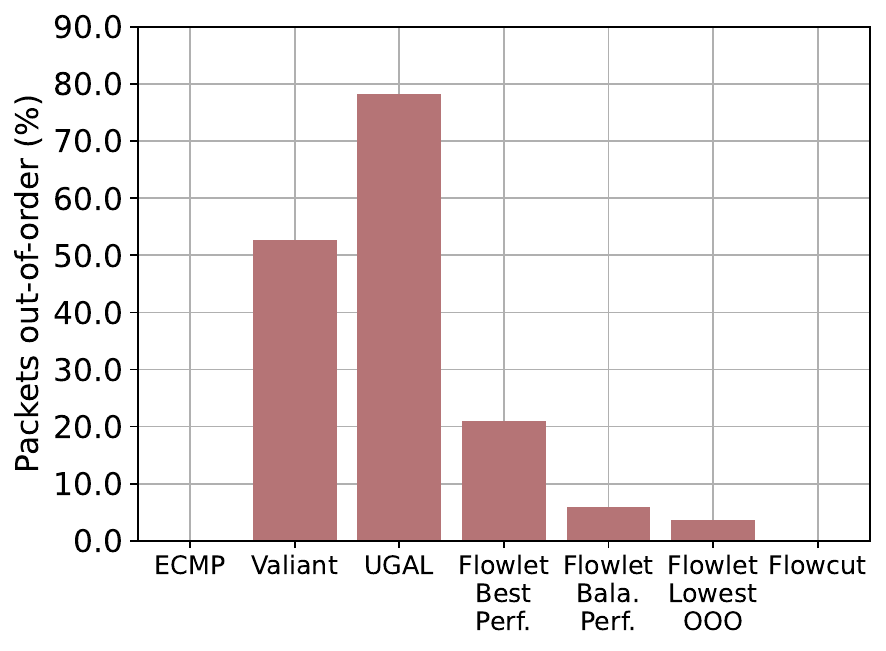}
\caption{Runtime and amount of out-of-order packets when using a random uniform distribution on Dragonfly.}
\label{fig:df1}
\end{figure}

We now analyze Flowcut performance on the Dragonfly topology, by comparing it with Flowlet switching, ECMP, UGAL~\cite{dragonfly}, and Valiant routing~\cite{valiant} (i.e., data is always forwarded through an intermediate random group). When using the Dragonfly topology, we can observe similar results, to those shown for fat trees. In particular, in Figure~\ref{fig:df1} we run the \textit{random uniform} distribution on the Dragonfly topology. The results show that Flowcut switching achieves performance close to Flowlet switching, and not too far behind UGAL, while guaranteeing in-order packet delivery. We also note how Valiant performs badly since it forces each packet to go through extra expensive hops, thus always increasing the latency and the overall completion time.

\begin{figure}[ht!]
\centering
\includegraphics[width=.5\columnwidth]{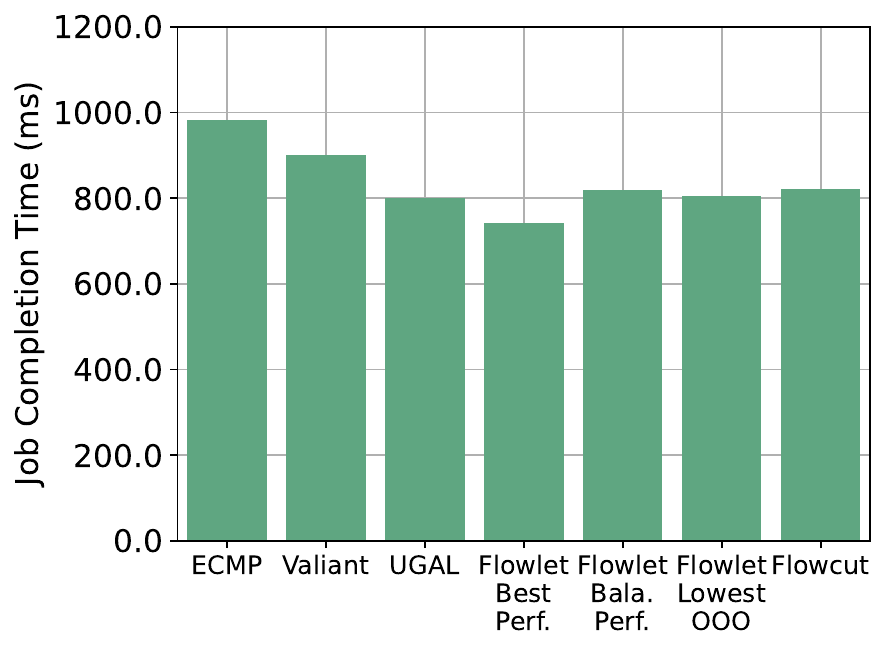}\hfill
\includegraphics[width=.5\columnwidth]{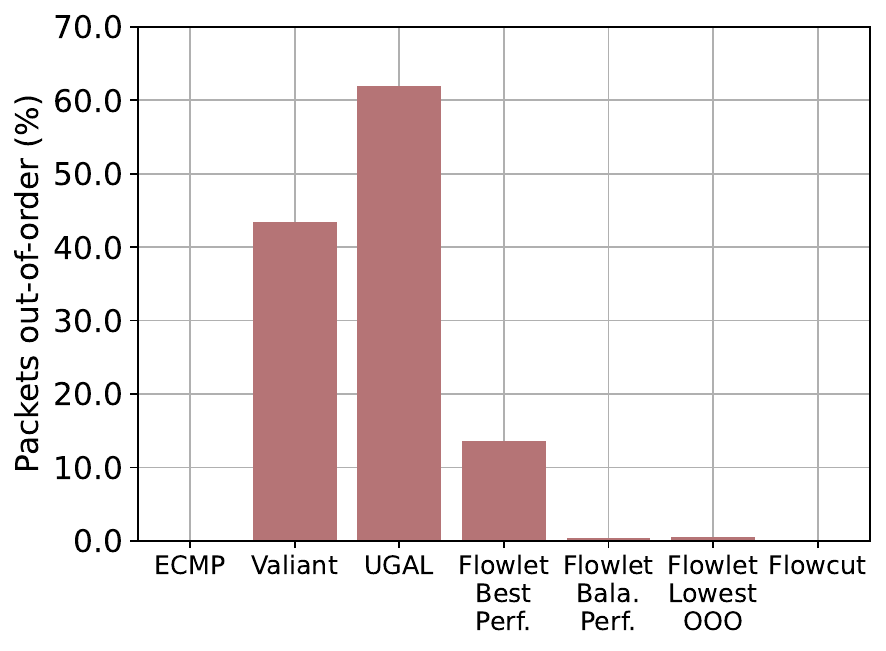}
\caption{Runtime and amount of out-of-order packets when using the enterprise distribution on Dragonfly.}
\label{fig:df2}
\end{figure}

A similar result can be observed in Figure~\ref{fig:df2} where we run the \textit{enterprise} distribution. In this case, there is no performance gap between Flowcut and UGAL, as well as between Flowcut and the balanced performing version of Flowlet. Like for the fat tree case with the data mining distribution, this is because the workload is characterized by many small flows, and hence routing has a smaller impact on FCT. Moreover, it seems changing path slightly less than on a per-packet basis is beneficial as it brings more stability and overall better performance.

\subsubsection{Empirical Results from a Slingshot System} \label{sec:real_hw_run}
To further evaluate the performance of Flowcut Switching, we conducted experiments on a Slingshot-based system that implements a variant of Flowcut Switching.

The experiments were performed on a Dragonfly topology consisting of 2,048 nodes, divided into 8 groups. Each group contains 16 switches, with 16 ports per switch dedicated to connecting to local NICs. The remaining ports are allocated for intra-group connections to other switches via local links and inter-group connections via global links. All links operate at a bandwidth of 200 Gbps.

We run an all-to-all traffic pattern and tested two routing modes of Slingshot. In the first, \textit{ordered} mode, packets are constrained to arrive in order at the receiver using Flowcut Switching. In the second, \textit{unordered} mode, this constraint is relaxed, resembling UGAL behavior as discussed in earlier examples. The \textit{ordered} mode is important for several applications and protocols, for example message envelopes and some amount of eager data, where the receiver needs to process the packets in a strict order to avoid encountering a large performance penalty ~\cite{10.1145/3405796.3405827,Ghasemirahni2022PacketOM}. On the other hand, several applications and protocols care less about such requirements and can have more relaxed constraints regarding the ordering of packets that they can exploit with Slingshot \textit{unordered} mode \cite{nvidia,multipath,ultraethernetUltraEthernet}.

\begin{figure}[ht!]
\centering
\includegraphics[width=\columnwidth]{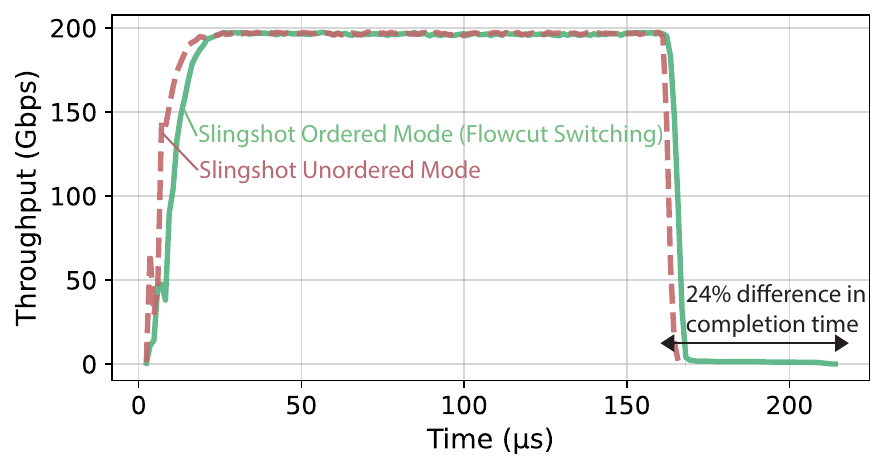}
\caption{Throughput of two Slingshot modes when running an all-to-all on a real Dragonfly system with 2,048 endpoints.}
\label{fig:real_data}
\end{figure}

The throughput results are presented in Figure~\ref{fig:real_data}. Notably, the ordered mode achieves throughput remarkably close to the unordered mode, with only a slight delay in reaching full utilization and a modest increase in completion time. Furthermore, both adaptive routing modes, ordered and unordered, outperform traditional non-adaptive systems such as Aries \cite{aries} by a significant margin based on internal data.
\section{Discussion}
\subsection{Packet Loss/Corruption}
\Name switching is designed to work with RoCE, that is usually deployed on lossless networks. If this is not the case, or if a packet gets discarded by the switch for any other reason (e.g., due to a CRC check failure), \name switching might not be able to resume paused flows. For example, this might happen if an ACK is lost after that the ingress switch paused a flow. In that case, the number of in-flight bytes will never become zero, and the flow will never be resumed. To avoid that, the source host keeps a timeout for each flow. If the flow is not resumed before the timeout expires, the source resumes the transmission. In that case, all subsequent packets will be routed by the switches on the old path. 


\subsection{Interaction with Congestion Control}\label{sec:design:discussion:cc}
Adaptive routing algorithms are orthogonal to congestion and flow control algorithms used by the hosts. In principle, the adaptive routing algorithm avoids most of the intermediate congestion~\cite{gpcnet,slingshot,10.1145/2674005.2674985}, preventing the congestion control algorithm to kick-in, by keeping the network bandwidth as high as possible. 
However, if congestion cannot be avoided through adaptive routing (as in the case of endpoint congestion generated by incast-like traffic~\cite{bfc,gpcnet}), the transmission rate of the flow at the source host will eventually decrease and match the bandwidth of the network path. 
For these reasons, we focused on non-incast scenarios in our evaluation since it is mostly a congestion control issue. Having said that, some of our evaluated scenarios (Figure~\ref{fig:df2}) do have some small incasts happening during their executions.


\subsection{Draining Time}\label{sec:discussion:drainingtime}
As discussed in Section~\ref{sec:flowchip:draining}, draining a flow introduces some overhead, because transmission of packets for that flow is stopped until the flow has no more in-flight packets. In general, this overhead is compensated because subsequent packets are sent on a less congested path. However, this depends on how many packets remain to be transmitted for that flow. To mitigate this issue, the source NIC could add an indication of the number of packets left for the flow, and the switches could drain the flow only if there are enough packets left, so that the cost of the draining will be compensated by the forwarding of the subsequent packets on a less congested path. Finally, we note that with certain protocols, even a single OOO packet might trigger re-transmissions which would always be more expensive (pending failure and edge cases) than draining the flow.

Alternatively, the switch could resume the transmission of the flow while there are still a few in-flight packets, to overlap the delivery of those packets to the destination host with the delivery of the resume packet to the source host. Although this would break the in-order delivery guarantee, it still guarantees a maximum \textit{Out-of-Order Degree} (OOD). The OOD is defined as the maximal difference between the sequence number of a received out-of-order packet, and the expected packet sequence number~\cite{8895760}, and some protocols can tolerate an OOD greater than zero. For example, in QUIC~\cite{quic} a retransmission is not triggered if the OOD is $\leq 3$, while other protocols tolerate an OOD $\leq 64$~\cite{8895760}. By allowing resuming the flow while there are still some in-flight packets we could reduce the draining time while providing guarantees on the maximum OOD. 

Regardless, even with our standard Flowcut Switching approach, we note that in our simulations the impact of draining is relatively small and greatly justifies the added performance of making the flow use multiple paths. We note this in Table~\ref{tab:drainingtime}, where we show the impact of draining for the fat tree experiments presented in Section~\ref{sec:evaluation:simulations}.

\begin{table}[]
\begin{tabular}{cc}
\textbf{Experiment}       & \textbf{\begin{tabular}[c]{@{}c@{}}Draining Impact \\ (avg. \% of the runtime spent draining)\end{tabular}} \\ \hline
Permutation               & 11.3\%                                                                                                       \\
Permutation with failures & 10.5\%                                                                                                       \\
Web Search & 5.2\%                                                                                                       \\
All-To-All                & 6.3\%                                                                                                       \\ \hline
\end{tabular}
\vspace{1pt} 
\caption{Showing the impact of draining on the total runtime. The number is the average considering all flows in the network.}
\label{tab:drainingtime}
\end{table}

\section{Related Work}
Several works address issues related to adaptive routing and in-order packet delivery. Some of these works balance traffic on a per flow granularity~\cite{ecmp,10.17487/RFC2992,hedera,vl2,10.1145/1397718.1397732,10.1145/2592798.2592803,10.1145/2535372.2535397,10.1145/2079296.2079304}. Although they guarantee in-order packet delivery, it has been shown that applications can still experience congestion due to the coarse granularity of the adaptive routing decisions~\cite{10.1145/2535372.2535375,conga,drill}. On the other hand, per packet adaptive routing algorithms react better to congestion~\cite{drill,10.1145/2535372.2535375,10.1145/2619239.2626309,10.1145/2342356.2342390,10.1145/2043164.2018467} but are characterized by higher number of OOO packets, posing significant challenges on the hosts for reordering packets. 
MPRDMA \cite{multipath} tries to balance this by pruning slow paths in order to reduce the amount of out-of-order packets. However, it cannot guarantee in-order delivery for all packets and requires ACK clocking to work.

Several adaptive routing algorithms rely on Flowlet switching to reduce the number of OOO packets~\cite{letitflow,conga,hula,10.1145/1232919.1232925}. However, they rely on the assumption that the traffic is bursty, that is usually not the case for RDMA-like protocols~\cite{8895760}. Moreover, they can only guarantee in-order delivery if the threshold used to detect flowlets is large enough. This depends on the network conditions, and having a too high threshold limits the adaptive routing opportunities and impacts the performance. For this reason, some algorithms dynamically tune this threshold according to the network traffic to balance performance and OOO packets~\cite{benet2019flowdyn,DIAO2022219,10.1145/1232919.1232925}, but they cannot still provide in-order delivery guarantees. 

ConWeave \cite{conweave} is a recently introduced solution that guarantees in-order delivery by temporarily buffering packets at the last switch before the receiver. Although it achieves impressive throughput and fairness, it has a few practical drawbacks compared to Flowcut: it requires substantial switch-side changes (while Flowcut can exist entirely on the NIC) and per-flow buffer queues (or traffic classes) at each ToR to hold rerouted packets; it depends on several microsecond-scale tunable timers (RTT-probe timeout, path-busy blacklist, tail-flush recovery); and its design and evaluation are tied to fat-tree topologies.

Other works implement the monitoring of congestion and adaptive routing decisions in the source host software stack~\cite{10.1145/2674005.2674985,259356}. For example, Flowbender~\cite{10.1145/2674005.2674985} monitors the congestion experienced by each flow by using \textit{Explicit Congestion Notification} (ECN) and, if a flow is too congested, it is rerouted on a different path. However, differently from \name switching, paths are selected randomly, and there is no guarantee that the newly selected path is less congested than the old one. Also, these approaches only take decisions at the endpoint, whereas \name switching can reroute packets at any point of the network, thus reacting more promptly to congestion.

Last, LEFT \cite{left} uses an endpoint based reorder buffer in a RNIC to absorb any out of order arrivals and deliver a perfectly in order stream to the application. This requires custom NIC support and allocates per flow buffers on the host.

\section{Conclusions}
In this paper we presented Flowcut switching, a simple and easy-to-implement mechanism to achieve good routing performance while guaranteeing in-order delivery of all packets. We show that Flowcut switching, unlike other state-of-the-art solutions like Flowlet switching,  requires minimal tuning and is more tolerant to parameter selection, independently of the workload, network configuration, congestion, and failures.
Most importantly, Flowcut switching can guarantee in-order delivery of packets. This is important for several protocols like RoCE, where packets are not handled efficiently if out-of-order. We remark that even a few out-of-order packets can cause problems to such protocols, if there is a large distance between their sequence numbers.
We propose three possible implementations for Flowcut switching, two requiring switch support (on all switches or only ingress switches) while another requiring only NIC changes.
In our testing using fat tree and dragonfly topologies, Flowcut switching consistently outperforms ECMP, by achieving 1.5x lower FCTs under normal conditions and by 5x when simulating link failures. Moreover, Flowcut switching also consistently outperforms a conservative version of Flowlet switching in all scenarios, as well as packet spraying in link failures scenarios.

\section{Acknowledgments}
This work is supported by the European
Union Commission’s Horizon Europe, under grant agreements
101175702 (NET4EXA), by the Sapienza University Grants ADAGIO
and D2QNeT (Bando per la ricerca di Ateneo 2023 and 2024), and by
the European Research Council (ERC) under the European Union’s
Horizon 2020 research and innovation program (grant agreement
PSAP, No. 101002047). We also thank the Swiss National Supercomputing Center (CSCS) for providing the computational resources used in this work.
The authors used ChatGPT and Claude solely for language editing and quality control; all ideas and content are original.

\bibliographystyle{IEEEtran}
\bibliography{bib}

\vfill

\end{document}